\documentclass[twocolumn, 12pt]{aastex7}
\newcommand{\Msun}{${\rm M}_\sun$}
\newcommand{\feh}{[Fe/H]}

\newcommand{\teff}{T$_{\rm eff}$}

\newcommand{\minesweeper}{\texttt{MINESweeper}}

\newcommand{\CocNUniqueRnd}{65}
\newcommand{\CocNDesiRnd}{48}
\newcommand{\CocNMageRnd}{23}
\newcommand{\CocNGaiaCandRnd}{279}
\newcommand{\CocNMageObsRnd}{62}
\newcommand{\CocFeHRnd}{-1.3}
\newcommand{\CocFeHMageFmt}{-1.33\pm0.04}
\newcommand{\CocFeHDesiFmt}{-1.24\pm0.02}
\newcommand{\CocFeHTableFmt}{-1.3\pm0.06}
\newcommand{\CocFeHDispMageFmt}{0.19\pm0.03}
\newcommand{\CocFeHDispDesiFmt}{0.17\pm0.02}
\newcommand{\CocFeHDispFmt}{0.18\pm0.02}
\newcommand{\CocFeHDispRnd}{0.2}
\newcommand{\CocSigmaVMageFmt}{6.5^{+1.2}_{-0.9}}
\newcommand{\CocSigmaVDesiFmt}{9.3^{+1.1}_{-0.9}}
\newcommand{\CocSigmaVFmt}{8.1\pm1.7}
\newcommand{\CocSigmaVRange}{6.5--9.3}
\newcommand{\CocWidthDegRnd}{1.6}
\newcommand{\CocWidthDegFmt}{1.6\pm0.2}
\newcommand{\CocWidthKpcFmt}{0.7\pm0.2}
\newcommand{\CocMvFmt}{-6.4\pm0.3}
\newcommand{\CocMstarFmt}{2.6\pm0.6}
\newcommand{\CocMvRelFmt}{-6.4\pm0.4}
\newcommand{\CocMstarRelFmt}{2.7\pm0.6}
\newcommand{\CocRARnd}{195}
\newcommand{\CocDecRnd}{-1}
\newcommand{\CocPMRAFmt}{1.1\pm0.2}
\newcommand{\CocPMDecFmt}{0.0\pm0.2}
\newcommand{\CocRVFmt}{55\pm9}
\newcommand{\CocDistFmt}{24\pm3}
\newcommand{\CocRrlDistRnd}{23}
\newcommand{\CocOrbitRARnd}{195}
\newcommand{\CocOrbitDecRnd}{-1}
\newcommand{\CocOrbitDistRnd}{25}
\newcommand{\CocOrbitPMRARnd}{1.1}
\newcommand{\CocOrbitPMDecRnd}{-0.03}
\newcommand{\CocOrbitRVRnd}{57}
\newcommand{\CocXRnd}{-0.7}
\newcommand{\CocYRnd}{-9.4}
\newcommand{\CocZRnd}{22.2}
\newcommand{\CocVxRnd}{142}
\newcommand{\CocVyRnd}{297}
\newcommand{\CocVzRnd}{51}
\newcommand{\CocApoRnd}{101}
\newcommand{\CocPeriRnd}{23}
\newcommand{\CocEccRnd}{0.6}
\newcommand{\CocLzFmt}{1.1\pm0.5}
\newcommand{\CocEtotFmt}{-0.5\pm0.1}
\newcommand{\CocDesiDistErrRnd}{2}
\newcommand{\CocDesiRVErrRnd}{1.2}
\newcommand{\CocMageDistErrRnd}{1.6}
\newcommand{\CocMageRVErrRnd}{0.5}

\usepackage{graphics}
\usepackage{amsmath}
\usepackage{longtable}
\usepackage{bm}
\graphicspath{ {./figures} }

\begin{document}

\title{The Cocytos Stream: A Disrupted Globular Cluster from our Last Major Merger?}

\author[0000-0003-2094-9128]{Christian~Aganze}
\affiliation{Kavli Institute for Particle Astrophysics \& Cosmology, Stanford University, Stanford, CA 94305, USA}
\email[show]{caganze@stanford.edu}

\author[0000-0002-0572-8012]{Vedant~Chandra}
\affiliation{Center for Astrophysics $\mid$ Harvard \& Smithsonian, 60 Garden St, Cambridge, MA 02138, USA}
\email{}

\author[0000-0003-2229-011X]{Risa~H.~Wechsler}
\affiliation{Kavli Institute for Particle Astrophysics \& Cosmology, Stanford University, Stanford, CA 94305, USA}
\email{}

\author[0000-0002-9110-6163]{Ting~S.~Li}
\affiliation{Department of Astronomy \& Astrophysics, University of Toronto, Toronto, ON M5S 3H4, Canada}
\email{}

\author[0000-0003-2644-135X]{Sergey~E.~Koposov}
\affiliation{Institute for Astronomy, University of Edinburgh, Royal Observatory, Blackford Hill, Edinburgh EH9 3HJ, UK}
\affiliation{Institute of Astronomy, University of Cambridge, Madingley Rd, Cambridge CB3 0HA, UK}
\email{}

\author[0000-0002-0740-1507]{Leandro {Beraldo~e~Silva}}
\affiliation{Department of Astronomy \& Steward Observatory, University of Arizona,
Tucson, AZ 85721, USA}
\affiliation{Observatório Nacional, Rio de Janeiro - RJ, 20921-400, Brasil}
\email{}

\author[0000-0002-5786-0787]{Andreia~Carrillo}
\affiliation{Institute for Computational Cosmology, Department of Physics, Durham University, Durham DH1 3LE, UK}
\email{}

\author[0000-0001-5805-5766]{Alexander~H.~Riley}
\affiliation{Institute for Computational Cosmology, Department of Physics, Durham University, Durham DH1 3LE, UK}
\email{}

\author[0000-0002-6257-2341]{Monica~Valluri}
\email{}
\affiliation{Department of Astronomy, University of Michigan, Ann Arbor, MI 48109, USA}

\author[0000-0001-9852-9954]{Oleg~Y.~Gnedin}
\affiliation{Department of Astronomy, University of Michigan, Ann Arbor, MI 48109, USA}
\email{}

\author[0000-0002-2446-8332]{Mairead~Heiger}
\affiliation{Department of Astronomy \& Astrophysics, University of Toronto, Toronto, ON M5S 3H4, Canada}
\email{mairead.heiger@astro.utoronto.ca}

\author[0000-0002-6667-7028]{Constance~Rockosi}
\affiliation{UC Observatories, University of California, Santa Cruz,  CA 95064, USA}
\email{}

\author[0000-0002-7667-0081]{Raymond~Carlberg}
\affiliation{Department of Astronomy \& Astrophysics, University of Toronto, Toronto, ON M5S 3H4, Canada}
\email{}

\author[0000-0002-5689-8791]{Amanda~Byström}
\affiliation{Institute for Astronomy, University of Edinburgh, Royal Observatory, Blackford Hill, Edinburgh EH9 3HJ, UK}
\email{}

\author[0000-0003-0853-8887]{Namitha~Kizhuprakkat}
\affiliation{Institute of Astronomy and Department of Physics, National Tsing Hua University, Hsinchu 30013, Taiwan}
\email{}

\author[0000-0002-2527-8899]{Mika~Lambert}
\affiliation{Department of Astronomy \& Astrophysics, University of California, Santa Cruz, Santa Cruz, CA 95064, USA}
\email{}

\author[0000-0002-8999-1108]{Bokyoung~Kim}
\affiliation{Institute for Astronomy, University of Edinburgh, Royal Observatory, Blackford Hill, Edinburgh EH9 3HJ, UK}
\email{}

\author[0000-0003-0105-9576]{Gustavo~Toledo~Medina}
\affiliation{Department of Astronomy \& Astrophysics, University of Toronto, Toronto, ON M5S 3H4, Canada}
\email{}

\author[0000-0002-0084-572X]{Carlos~Allende~Prieto}
\affiliation{Departamento de Astrof\'{\i}sica, Universidad de La Laguna (ULL), E-38206, La Laguna, Tenerife, Spain}
\affiliation{Instituto de Astrof\'{\i}sica de Canarias, C/ V\'{\i}a L\'{a}ctea, s/n, E-38205 La Laguna, Tenerife, Spain}
\email{}

\author{Jessica~Nicole~Aguilar}
\affiliation{Lawrence Berkeley National Laboratory, 1 Cyclotron Road, Berkeley, CA 94720, USA}
\email{}

\author[0000-0001-6098-7247]{Steven~Ahlen}
\affiliation{Physics Dept., Boston University, 590 Commonwealth Avenue, Boston, MA 02215, USA}
\email{}

\author[0000-0001-9712-0006]{Davide~Bianchi}
\affiliation{Dipartimento di Fisica ``Aldo Pontremoli'', Universit\`a degli Studi di Milano, Via Celoria 16, I-20133 Milano, Italy}
\affiliation{INAF-Osservatorio Astronomico di Brera, Via Brera 28, 20122 Milano, Italy}
\email{}

\author{David~Brooks}
\affiliation{Department of Physics \& Astronomy, University College London, Gower Street, London, WC1E 6BT, UK}
\email{}

\author{Todd~T.~Claybaugh}
\affiliation{Lawrence Berkeley National Laboratory, 1 Cyclotron Road, Berkeley, CA 94720, USA}
\email{}

\author[0000-0001-8274-158X]{Andrew~P.~Cooper}
\affiliation{Institute of Astronomy and Department of Physics, National Tsing Hua University, 101 Kuang-Fu Rd. Sec. 2, Hsinchu 30013, Taiwan}
\email{}

\author[0000-0002-1769-1640]{Axel~de~la~Macorra}
\affiliation{Instituto de F\'{\i}sica, Universidad Nacional Aut\'{o}noma de M\'{e}xico,  Circuito de la Investigaci\'{o}n Cient\'{\i}fica, Ciudad Universitaria, Cd. de M\'{e}xico  C.~P.~04510,  M\'{e}xico}
\email{}

\author[0000-0002-4928-4003]{Arjun~Dey}
\affiliation{NSF NOIRLab, 950 N. Cherry Ave., Tucson, AZ 85719, USA}
\email{}

\author{Peter~Doel}
\affiliation{Department of Physics \& Astronomy, University College London, Gower Street, London, WC1E 6BT, UK}
\email{}

\author[0000-0002-2890-3725]{Jaime~E.~Forero-Romero}
\affiliation{Departamento de F\'isica, Universidad de los Andes, Cra. 1 No. 18A-10, Edificio Ip, CP 111711, Bogot\'a, Colombia}
\affiliation{Observatorio Astron\'omico, Universidad de los Andes, Cra. 1 No. 18A-10, Edificio H, CP 111711 Bogot\'a, Colombia}
\email{}

\author{Enrique~Gaztañaga}
\affiliation{Institut d'Estudis Espacials de Catalunya (IEEC), c/ Esteve Terradas 1, Edifici RDIT, Campus PMT-UPC, 08860 Castelldefels, Spain}
\affiliation{Institute of Cosmology and Gravitation, University of Portsmouth, Dennis Sciama Building, Portsmouth, PO1 3FX, UK}
\affiliation{Institute of Space Sciences, ICE-CSIC, Campus UAB, Carrer de Can Magrans s/n, 08913 Bellaterra, Barcelona, Spain}
\email{}

\author[0000-0003-3142-233X]{Satya~A.~Gontcho}
\affiliation{Lawrence Berkeley National Laboratory, 1 Cyclotron Road, Berkeley, CA 94720, USA}
\email{}

\author{Gaston~Gutierrez}
\affiliation{Fermi National Accelerator Laboratory, PO Box 500, Batavia, IL 60510, USA}
\email{}

\author[0000-0002-6024-466X]{Mustapha~Ishak}
\affiliation{Department of Physics, The University of Texas at Dallas, 800 W. Campbell Rd., Richardson, TX 75080, USA}
\email{}

\author[0000-0003-3510-7134]{Theodore~Kisner}
\affiliation{Lawrence Berkeley National Laboratory, 1 Cyclotron Road, Berkeley, CA 94720, USA}
\email{}

\author[0000-0001-6356-7424]{Anthony~Kremin}
\affiliation{Lawrence Berkeley National Laboratory, 1 Cyclotron Road, Berkeley, CA 94720, USA}
\email{}

\author{Ofer~Lahav}
\affiliation{Department of Physics \& Astronomy, University College London, Gower Street, London, WC1E 6BT, UK}
\email{}

\author[0000-0003-1838-8528]{Martin~Landriau}
\affiliation{Lawrence Berkeley National Laboratory, 1 Cyclotron Road, Berkeley, CA 94720, USA}
\email{}

\author[0000-0001-7178-8868]{Laurent~Le~Guillou}
\affiliation{Sorbonne Universit\'{e}, CNRS/IN2P3, Laboratoire de Physique Nucl\'{e}aire et de Hautes Energies (LPNHE), FR-75005 Paris, France}
\email{}

\author[0000-0002-1125-7384]{Aaron~Meisner}
\affiliation{NSF NOIRLab, 950 N. Cherry Ave., Tucson, AZ 85719, USA}
\email{}

\author{Ramon~Miquel}
\affiliation{Instituci\'{o} Catalana de Recerca i Estudis Avan\c{c}ats, Passeig de Llu\'{\i}s Companys, 23, 08010 Barcelona, Spain}
\affiliation{Institut de F\'{i}sica d’Altes Energies (IFAE), The Barcelona Institute of Science and Technology, Edifici Cn, Campus UAB, 08193, Bellaterra (Barcelona), Spain}
\email{}

\author[0000-0003-3188-784X]{Nathalie~Palanque-Delabrouille}
\affiliation{IRFU, CEA, Universit\'{e} Paris-Saclay, F-91191 Gif-sur-Yvette, France}
\affiliation{Lawrence Berkeley National Laboratory, 1 Cyclotron Road, Berkeley, CA 94720, USA}
\email{}

\author[0000-0001-7145-8674]{Francisco~Prada}
\affiliation{Instituto de Astrof\'{i}sica de Andaluc\'{i}a (CSIC), Glorieta de la Astronom\'{i}a, s/n, E-18008 Granada, Spain}
\email{}

\author[0000-0001-6979-0125]{Ignasi~Pérez-Ràfols}
\affiliation{Departament de F\'isica, EEBE, Universitat Polit\`ecnica de Catalunya, c/Eduard Maristany 10, 08930 Barcelona, Spain}
\email{}

\author{Graziano~Rossi}
\affiliation{Department of Physics and Astronomy, Sejong University, 209 Neungdong-ro, Gwangjin-gu, Seoul 05006, Republic of Korea}
\email{}

\author[0000-0002-9646-8198]{Eusebio~Sanchez}
\affiliation{CIEMAT, Avenida Complutense 40, E-28040 Madrid, Spain}
\email{}

\author{Michael~Schubnell}
\affiliation{Department of Physics, University of Michigan, 450 Church Street, Ann Arbor, MI 48109, USA}
\email{}

\author{David~Sprayberry}
\affiliation{NSF NOIRLab, 950 N. Cherry Ave., Tucson, AZ 85719, USA}
\email{}

\author[0000-0003-1704-0781]{Gregory~Tarl\'{e}}
\affiliation{University of Michigan, 500 S. State Street, Ann Arbor, MI 48109, USA}
\email{}

\author{Benjamin~A.~Weaver}
\affiliation{NSF NOIRLab, 950 N. Cherry Ave., Tucson, AZ 85719, USA}
\email{}

\author[0000-0002-6684-3997]{Hu~Zou}
\affiliation{National Astronomical Observatories, Chinese Academy of Sciences, A20 Datun Road, Chaoyang District, Beijing, 100101, P.~R.~China}
\email{}

\begin{abstract}

\noindent
The census of stellar streams and dwarf galaxies in the Milky Way provides direct constraints on galaxy formation models and the nature of dark matter. 
The DESI Milky Way survey --- with a footprint of $14,000$~deg$^2$ and a depth of $r<19$~mag --- delivers the largest sample of distant metal-poor stars compared to previous optical fiber-fed spectroscopic surveys.  
This makes DESI an ideal survey to search for previously undetected streams and dwarf galaxies. 
We present a detailed characterization of the Cocytos stream, which was re-discovered using a clustering analysis with a catalog of giants in the DESI year 3 data, supplemented with Magellan/MagE spectroscopy. 
Our analysis reveals a relatively metal-rich ([Fe/H]\,$=\CocFeHRnd$) and thick stream (width\,$=\CocWidthDegRnd^\circ$) at a heliocentric distance of $\approx \CocOrbitDistRnd$~kpc, with an internal velocity dispersion of \CocSigmaVRange~km\,s$^{-1}$. 
The stream's metallicity, radial orbit, and proximity to the Virgo stellar overdensities suggest that it is most likely a disrupted globular cluster that came in with the Gaia--Enceladus merger. We also confirm its association with the Pyxis globular cluster. Our result showcases the ability of wide-field spectroscopic surveys to kinematically discover faint disrupted dwarfs and clusters, enabling constraints on the dark matter distribution in the Milky Way.

\end{abstract}

\section{Introduction}

In the current paradigm of hierarchical galaxy formation models with dark matter, the accretion of low-mass halos plays an essential role in the formation of galaxies \citep{1978MNRAS.183..341W, 1999MNRAS.307..495H,2005ApJ...635..931B}. Therefore, probing accreted debris in the stellar halos of massive galaxies directly tests these models. The Milky Way halo is an essential laboratory for testing different galaxy formation frameworks because of its proximity, allowing us to measure its chemical composition and dynamics in detail. The era of large-scale photometric, spectroscopic, and astrometric surveys (e.g., LAMOST, \citealt{2012RAA....12.1197C}; DES, \citealt{2016MNRAS.460.1270D}; Pan-STARRS, \citealt{2016arXiv161205560C}; SDSS, \citealt{2017AJ....154...28B}; H3, \citealt{2019ApJ...883..107C}; GALAH, \citealt{2021MNRAS.506..150B}; Gaia, \citealt{2023A&A...674A...1G}) has begun to untangle the complex formation and evolution history of the Galaxy. In addition to probing the internal chemodynamics of the Milky Way, we have now mapped the outskirts of the Galaxy and discovered many Milky Way satellites. Together, samples of Milky Way satellites and precise chemodynamical maps of structures embedded in the Galactic halo provide new insights into the structure, formation, and evolution of the Galaxy. For example, Gaia has revealed that the Galactic halo is primarily composed of debris from its past merger history (e.g., the Gaia--Enceladus, Sequoia, and Helmi streams; \citealt{2018MNRAS.478..611B,2018Natur.563...85H,2018ApJ...856L..26M,2020ARA&A..58..205H,2020ApJ...901...48N}). Moreover, these surveys have uncovered $\approx$60 satellite galaxies which provide constraints on small-scale differences in dark-matter models \citep{2007ApJ...670..313S,2020ApJ...893...47D, 2020ApJ...890..136M,2021PhRvL.126i1101N, 2022ApJ...935L..17S,2023ApJ...953....1C,2025ApJ...979..176T}; the orbits of these satellites help build a mass model for the Galaxy \citep{2019MNRAS.484.5453C}, and they provide constraints on the limits of galaxy formation \citep{2021PhRvL.126i1101N}. 

One significant contribution of these large-scale Milky Way surveys is the discovery and characterization of stellar streams, which the Gaia data have revolutionized. Stellar streams form through tidal disruption of satellite galaxies and globular clusters by a massive host galaxy. They retain information about the orbital properties of their progenitors even when they are phase-mixed, and have features that persist on timescales of millions to billions of years \citep{2000MNRAS.319..657H}. As streams trace the orbit of their progenitor, they have been used to constrain the mass model of the Galaxy \citep{1999ApJ...512L.109J,2004ApJ...610L..97H,2010ApJ...712..260K,2016ApJ...833...31B,2019MNRAS.486.2995M,2023arXiv231117202I, 2023MNRAS.521.4936K} and constrain the dynamical interaction between the Milky Way and its satellites \citep{2021MNRAS.501.2279V}. Data from large-scale surveys, supplemented by recent precise astrometry from Gaia, have been the main drivers in the discovery and characterization of stellar streams in the Galaxy. Roughly $\approx$~100 streams have now been discovered in the Milky Way, with a significant fraction dominated by streams with globular cluster progenitors \citep{2023MNRAS.tmp..322M,2024arXiv240519410B}. Nonetheless, work is needed to understand the completeness of current surveys and to match the observed distribution of streams to predictions from cosmological models \citep{2022ApJ...928...30L, 2023ApJ...949...44S, 2024arXiv240515851P,2024arXiv241009143S,2024arXiv241009144R,2024arXiv240913810D}. There are hints that current samples may still be incomplete.

Although photometric and astrometric searches tend to cover large areas of the sky, they are shallower and less sensitive than searches that incorporate line-of-sight velocity information. Furthermore, a detailed chemical analysis of these streams, which is needed to constrain the nature of their progenitors, requires spectroscopic follow-up. Using a 6D kinematic search in the H3 Survey \citep{2019ApJ...883..107C}, \cite{2022ApJ...940..127C} discovered an extremely faint disrupting dwarf galaxy that they named {\it Specter}. It has a stellar mass of $\approx 2000$\,\Msun~and \feh$=-1.8$, and is located at a distance of $\approx 12$~kpc. They predict that dozens of similar faint objects are yet to be detected in the Galaxy, which should be identifiable in large-area spectroscopic surveys. In this work, we conducted a similar search for previously undetected streams and disrupting dwarfs with strong clustering in line-of-sight velocities in the DESI survey.

The Dark Energy Spectroscopic Instrument (DESI, \citealt{2016arXiv161100036D,2022AJ....164..207D}) is a fiber-fed spectrograph mounted on the 4-meter Mayall telescope at the Kitt Peak National Observatory, covering wavelength ranges of $\approx 0.4-1\,\micron$ at a spectral resolution $(\lambda/ \Delta \lambda)$ of $2500$-$5000$. DESI has a field of view of $\approx$3 degrees, allowing it to obtain simultaneous spectra of $\approx$5,000 objects per pointing. The full DESI survey aims to obtain precise redshifts for tens of millions of galaxies, quasars, and stars. As part of the bright time survey, the DESI Milky Way Survey (MWS, \citealt{2023ApJ...947...37C}) aims to obtain spectra of $> 10$~million stars over five years, down to a magnitude limit of $r \approx 19$, covering a sky area of $\approx 14,000$~degree$^2$. This survey will yield the largest number of distant ($>10$~kpc) and metal-poor stars ([Fe/H]$<-1.5$) in the halo compared to any existing spectroscopic survey \citep{2023ApJ...947...37C,2024MNRAS.533.1012K,2024arXiv240206242Z}. Additionally, the planned DESI MWS footprint includes $\approx 31$~dwarf galaxies and $\approx40$~known stellar streams, with a minimum of $10$-$100$ members per stream, allowing us to characterize known structures in detail.

This paper presents the kinematic discovery of a stellar stream identified with a blind search in the first year of DESI data and further characterization with DESI year 3 data and additional Magellan/MagE follow-up. This stream coincides with the Cocytos stream that was discovered photometrically by \cite{2009ApJ...693.1118G} in the SDSS data, using a matched-filter isochrone search technique, with a photometric distance of $\approx$~11~kpc and a width of 0.7~$^\circ$. Based on this width, they postulated that it likely had a GC progenitor. Four likely member stars of Cocytos have also been detected in Gaia and have been associated with the Virgo Overdensity (VOD, \citealt{2008ApJ...673..864J,2019ApJ...886...76D}). The VOD has been proposed to be the outer shells of the Gaia--Enceladus merger. We explore these scenarios in more detail here. Section \ref{sec:data} describes the data and follow-up observations, Section \ref{sec:methods}
describes our analysis of the stream, and Sections \ref{sec:discuss} and \ref{sec:summary} discuss and summarize our results.

\section{Data and Stream Discovery}\label{sec:data}

\subsection{The DESI Milky Way Survey and Data Products}

The target selection, observation strategy, and data reduction and analysis pipelines of the DESI Milky Way Survey (MWS) are summarized by \cite{2020RNAAS...4..188A,2023ApJ...947...37C,2024MNRAS.533.1012K,2024arXiv241009149B}. Briefly, the MWS selects targets from the Gaia DR3 catalog and the DESI Legacy Imaging Surveys \citep{2019AJ....157..168D} for subsequent spectroscopic follow-up. The survey is optimized to obtain chemical abundances and radial velocities for stars with Gaia $G >16$~mag, beyond the nominal limit for Gaia radial velocity targets, and collect the largest number of spectra compared to previous spectral surveys of the Galaxy (e.g. RAVE, H3, LAMOST). To define targeting prioritization, the DESI MWS target sample is divided into primary targets, secondary targets (spare fiber program), and backup targets (Dey et al., in preparation). Each program is further divided into main ($r<19$), faint ($r>19$), and high-priority targets, which generally consist of white dwarfs, RR Lyrae (RRLs), nearby ($<100$~pc) stars, and blue horizontal branch (BHB) stars. Additionally, the DESI MWS also observes known dwarf galaxies and globular clusters as secondary targets along with other additional targets, and the backup program observes disk stars and bright halo stars ($10.5<G<16$).

The DESI MWS strategy, fiber assignment, and prioritization are constrained by the main DESI galaxy survey, which has priority over MWS targets. The MWS (and other smaller surveys) use the time left over from the main survey, which allows it to maximize the observing time of the entire DESI survey. In the MWS, targets are generally exposed on the fly depending on prioritization to reach an effective exposure time of $\approx$ 180~seconds to obtain a large number of adequate S/N spectra per night. DESI MWS spectra are processed through many pipelines; we used stellar parameters from the latest DESI internal Loa release (12th) through the RVS pipeline, which will be made public with DESI DR2 (year 3 data). 

The RVS pipeline uses the {\tt RVSpecfit } code \footnote{\url{https://github.com/segasai/rvspecfit}} \citep{2019MNRAS.485.4726K} to fit synthetic spectra from the PHOENIX grid \citep{2013A&A...553A...6H} to co-added DESI spectra using Nelder--Mead optimization \citep{neldermead65}.  {\tt RVSpecfit } determines uncertainties on best-fit parameters using the covariance matrix outputted by the optimization algorithm.

\subsection{Distance Estimation}

We used spectrophotometric distances obtained using the {\tt rvsdistnn} method by Koposov et al. (in prep.). Briefly,  {\tt rvsdistnn} implements a neural network to map stellar parameters $\log g$, $T_{\rm eff}$, [Fe/H], $[\alpha/Fe]$ derived by the {\tt RVSpecFit} pipeline \citep{2019ascl.soft07013K} and stellar colors into absolute magnitude:
\begin{equation}
    M_{X} = F(\log g, T_{\rm eff}, {\rm [Fe/H]}, {\rm[}\alpha/{\rm Fe]} , C),
\end{equation}
where $F$ is the fully connected neural network with three hidden layers. Each hidden layer has $64$ inputs/outputs with a rectified linear unit (RELU) activation function. Two networks were trained, one predicting the Gaia G-band magnitude and using the Gaia BP-RP color, another predicting the absolute magnitude in Legacy Survey $r$ and using the $g-r$ color. The networks were trained on the whole set of 15 million {\tt RVSpecFit} measurements from DESI spectra that will be released as part of the DESI DR2 by optimizing the loss function defined as:
\begin{align}
\mathrm{Loss} = \sum_i 
\left(\frac{{\omega_i}  - 10^{ 0.2 \, F_i -0.2 \, m_i+ 2} }{\sigma_{\omega,i}}\right)^2,     
\end{align}
where index $i$ refers to the $i$-th star in the training data, $F_i $ is the application of the neural network on the stellar parameters of the $i$-th star $ F(\log g_i, T_{\rm eff, i}, {\rm [Fe/H]}_i, {\rm[}\alpha/{\rm Fe]}_i, C_i)$, $m_i$ is apparent extinction corrected magnitude, and $\omega_i$,$\sigma_{\omega,i}$ are Gaia parallaxes and uncertainties in mas with corrections from \cite{2021A&A...649A...2L}. This method implemented the neural network using the {\tt pytorch} framework, and we trained it using stochastic gradient descent with the {\tt Adam} optimizer and iterative reduction of the learning rate when encountering a plateau in the loss function. To estimate the uncertainty in predicted absolute magnitudes, this method Monte-Carlo-sampled the stellar parameter uncertainties (100 times) and propagated them through the network. In the case of stars with Gaia and Legacy Survey photometry, the absolute magnitudes with the smallest uncertainty were picked. The distance moduli for each star were then computed from the apparent and predicted absolute magnitudes.

\subsection{Clustering Analysis with DESI Giants}

We selected giant stars in DESI using their Gaia DR3 astrophysical parameters with $\log g  < 3.5$, 3000~K $<$ \teff $<$6000~K and with parallaxes $<1$ mas. To search for distant clusters, we focused on structures in the outer halo by selecting 33,944 giants with distances  $>10$~kpc. We used the density-based Ordering Points to Identify the Clustering Structure (OPTICS) algorithm \citep{ankerst1999optics}. OPTICS is a clustering method designed to identify clusters of varying densities by ordering data points based on their distances. We used the \texttt{scikit-learn} implementation of OPTICS, which only requires the minimum number of points in a cluster as an input parameter. Many methods have been used to identify streams and substructures in the Milky Way, but there is no single best-performing stream-finding method; a brief review is given by \cite{2024arXiv240105620W}. Examples include a pole-counting technique by \cite{1996ApJ...465..278J}, which identifies stellar overdensities along particular great circles; matched-filter isochrone searches \citep{2006ApJ...639L..17G,2009ApJ...693.1118G,2018ApJ...862..114S}; self-organizing maps \citep{2018ApJ...863...26Y}; wavelet transforms \citep{2012MNRAS.426L...1A, 2021ApJ...915...48D}; density-based clustering \citep{2023A&A...670L...2D}, deep-learning-based clustering \citep{2024arXiv240105620W}, anomaly detection methods \citep{2022MNRAS.509.5992S,2024MNRAS.527.8459P}, and the identification of streams based on their orbital tracks assuming a Galactic potential \cite[\texttt{STREAMFINDER}]{2018MNRAS.477.4063M}. 

We masked two major streams: the Sagittarius stream based on the track by \cite{2020A&A...635L...3A}, and the Orphan--Chenab stream based on the track by \cite{2023MNRAS.521.4936K}, discarding any giants within 15$^\circ$ of the streams. Our clustering analysis used on-sky coordinates, proper motions, distances, and radial velocities as clustering features. To determine clusters, we used on-sky positions, proper motions, distances and radial velocities separately re-scaled by their variance to obtain a dimensionless renormalization \citep{2011ApJ...738...79X}. We then implemented OPTICS, fixing the minimum number of stars that define a cluster to the lowest possible number of three stars per cluster. We used the Euclidean distance metric as an estimator for distances between points. 

We identified 41 clusters with radial velocity dispersions $<25$ km s$^{-1}$ in DESI year 1 data. As a next step, we compared the distribution of all clusters to known structures (streams, globular clusters, dwarf galaxies, and other halo substructures), primarily by comparing their sky positions with the new clusters within a sky separation of $5$$^\circ$. We also checked these structures against tracks of known stellar streams in the \texttt{Galstreams} catalog \citep{2023MNRAS.tmp..322M}, globular clusters in the \cite{2021MNRAS.505.5957B} catalog and Local Volume Database of known dwarf galaxies \citep{2024arXiv241107424P}\footnote{ We used the catalog provided at \url{https://raw.githubusercontent.com/apace7/local_volume_database/main/data/dwarf_mw.csv}}.

\subsection{Searching for Gaia Giant Members with Isochrone and Proper Motion Selections}\label{sec:gaiaselection}

Most clusters we detected at distances $>30$~kpc were near known streams or dwarf galaxies. However, several known streams have no reported radial velocity measurements in the literature (see \citealt{2024arXiv240519410B}). In this context, we further investigated a cluster of $5$ giant members in the DESI year 1 data, with median $(\alpha, \delta) =(183, -5) $~deg. and median  $(\mu_{\alpha}, \mu_{\delta }*) = (1.3, -0.1) $~$\mathrm{mas~yr}^{-1}$. This cluster is located near Cocytos on the sky, which thus far has only been identified in SDSS with photometry by \cite{2009ApJ...693.1118G}. Based on this initial cluster of five giants, we further searched for more members in Gaia.
We queried sources in Gaia DR3 in its vicinity, requiring distances $>10$~ kpc and masking the Galactic plane as: 

{\small
\begin{verbatim}
SELECT *, DISTANCE(183.25, -5.23, ra, dec) AS ang_sep
FROM gaiadr3.gaia_source
WHERE DISTANCE(183.25, -5.23, ra, dec) < 85 
AND Abs(b)>20
AND NOT parallax >0.1
AND pmra >-3
AND pmra <3
AND pmdec >-1
AND pmdec <1
AND phot_g_mean_mag <21
\end{verbatim}
}

This process resulted in $\approx$1.6 million stars. We then applied an isochrone and proper-motion-based selection to the Gaia stars centered around the DESI giants to further select members. We first obtained a grid of pre-computed isochrones from the MESA Isochrones and Stellar Tracks (MIST) grid \citep{2016ApJ...823..102C,2016ApJS..222....8D} version 1.2, using the \texttt{artpop} package \citep{2022ApJ...941...26G}. We selected the red giant track by requiring the equivalent evolutionary point (EEP) parameter provided by the isochrone to be $\in$ [400, 800] and $\log$g$< 4$. As a next step, we applied reddening correction to the Gaia BP, RP, and G magnitudes of Gaia stars by adopting E(B-V) using the 2D SFD reddening maps \citep{1998ApJ...500..525S} with the \texttt{dustmaps} package \citep{2018JOSS....3..695M}, and reddening coefficients by \cite{2019ApJ...877..116W}. We assumed a cluster distance of \CocOrbitDistRnd~kpc, the inverse-variance-weighted orbit distance of the combined DESI and MagE members. 
Lastly, we selected members by requiring a broad metallicity selection consisting of four different metallicities ([Fe/H]=$-1, \CocFeHRnd, -1.5 {\rm ~and} -2$) and selected giants with  $| {\rm G}-(G_{\rm iso} \pm  {\rm G}_{\rm err})| < 0.5 {\rm ~mag} $. ${\rm G}_{\rm iso}$ is obtained by interpolating the isochrone grid in BP-RP vs G~mag space.

\begin{figure*}
    \centering
    \includegraphics[width=\textwidth]{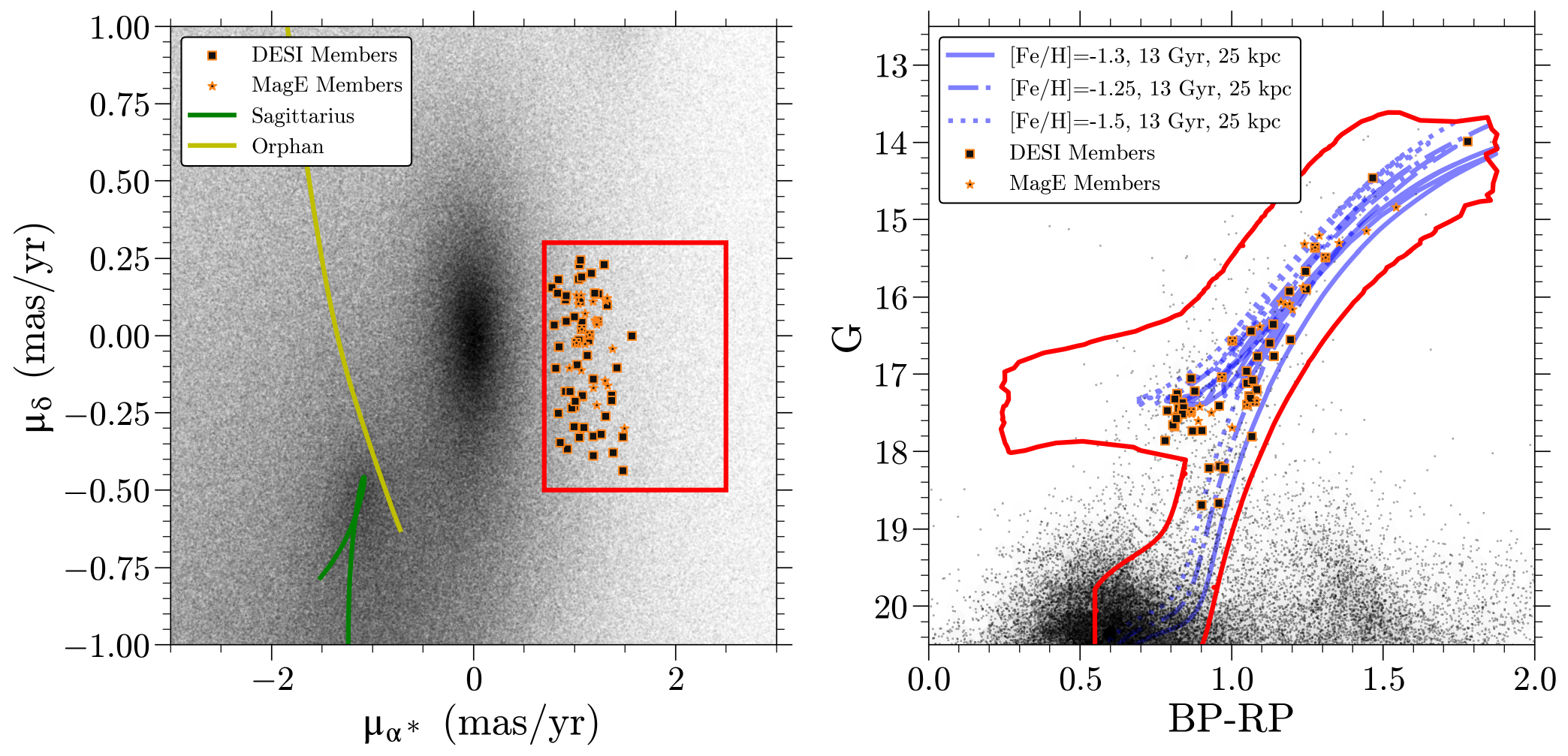}
    \caption{Illustration of the isochrone and proper-motion selection applied to the Gaia catalog within $|\phi_1|<50$$^\circ$ and $\phi_2<20 $$^\circ$ in stream coordinates, defined from two poles: (R.A., dec)= (180, -10)$^\circ$ and (230, 17)$^\circ$.  {\em Left:} Distribution of proper motions for all stars in the Gaia query, with our selection for Cocytos shown as a red rectangle. Black rectangles and stars with orange contours show our final radial-velocity-selected members with DESI and MagE (this spectroscopic selection is discussed in the main text). Two lines show the tracks of the Sagittarius and Orphan streams within $|\phi_1|<50$$^\circ$ and $\phi_2<20 $$^\circ$ of the stream. {\em Right:}  We show our isochrone selection using a reddening-corrected Gaia color--magnitude diagram with a red line. Blue lines show MIST isochrones for an age of 13 Gyr with three different metallicities, offset to the distance of \CocOrbitDistRnd~kpc. Spectroscopic members of Cocytos are shown as black rectangles and stars with orange contours. This proper motion selection shows an overdensity along the isochrone tracks, which is also in agreement with spectroscopic members.  }
    \label{fig:pmselection}
\end{figure*}

\begin{figure*}
    \centering
    \includegraphics[width=\textwidth]{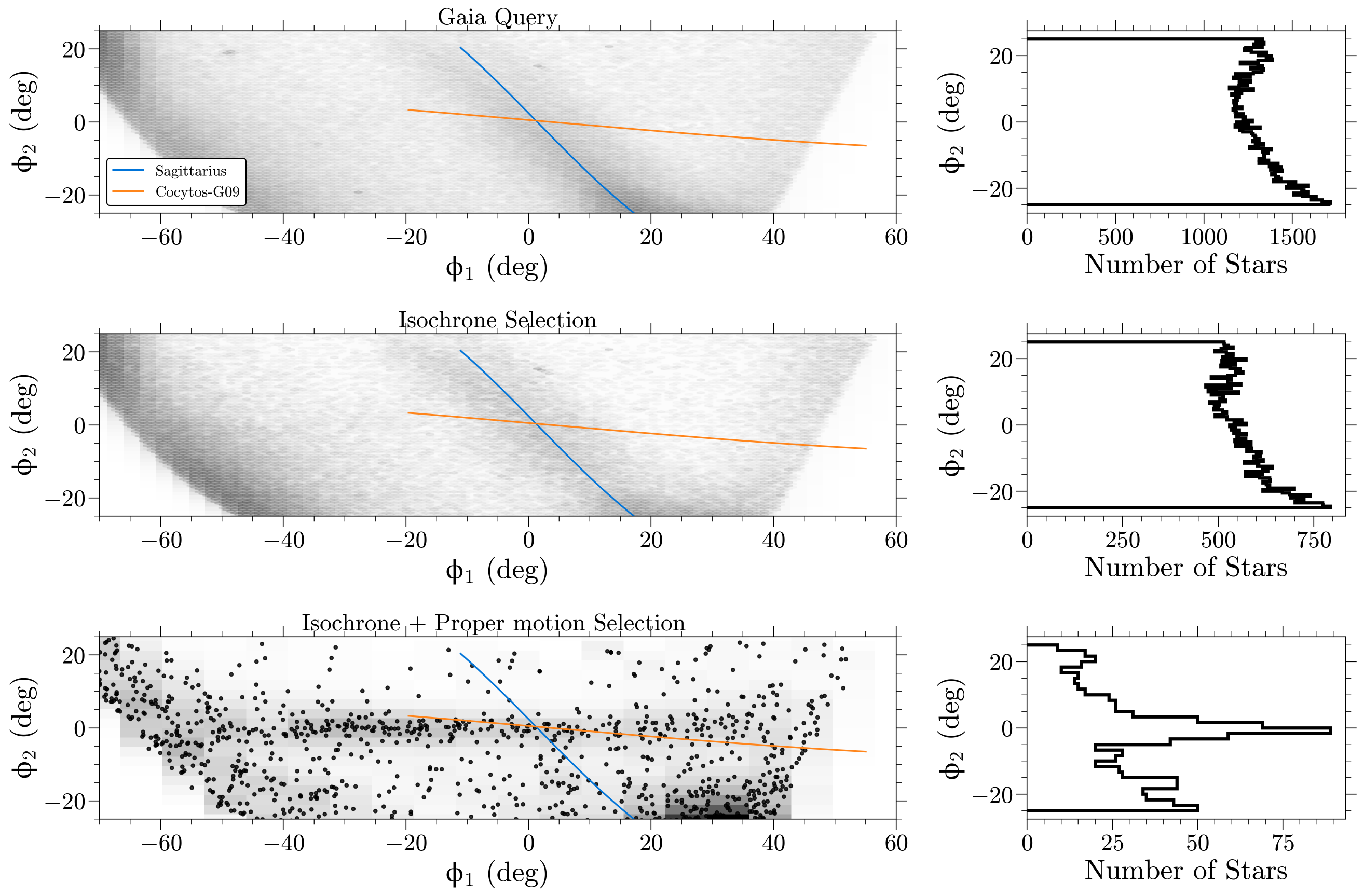}
    \caption{Selection of Cocytos in Gaia visualizing the stellar density along Great-circle coordinates of the stream. The top panels show the density distribution of stars in the Gaia query down to a limit of G=19. Middle panels show the isochrone-selected population for [Fe/H] = -2, -1.5 and -1. The bottom panels show the distribution of stars as a histogram and as scatter plots. Our selection shows a peak in the stellar density along $\phi_2=0 $ deg. }
    \label{fig:gaiaselection}
\end{figure*}

Based on the initial sets of DESI giants, we defined a simple proper-motion selection that required $ \mu_{\alpha*} \in [0.7, 2.5]$~ $\mathrm{mas~yr}^{-1}$ and $ \mu_{\delta} \in [-0.5, 0.3]$~$\mathrm{mas~yr}^{-1}$. This selection ensured that we included most DESI giants and a significant fraction of their orbits while avoiding contamination from the broader Milky Way halo and nearby streams. Furthermore, this simple selection in proper motion resulted in an identifiable overdensity along the isochrone tracks (Figure \ref{fig:pmselection}). Figure \ref{fig:gaiaselection} shows the stellar overdensity on the sky that resulted from the combination of proper motion and isochrone masks. Our isochrone selection at this distance and metallicity range shows significant contamination from the Sagittarius stream, but the proper motion constraints remove this contamination, showing a clear stream near $\phi_2=0$. Within $|\phi_1|<50$~$^\circ$ and $|\phi_2|<5$~$^\circ$, we obtained \CocNGaiaCandRnd{} candidate giant members at the $G=19$ limit. The stream extends to nearly $\approx$~80$^\circ$, making it nearly as extended as other prominent streams (e.g., Cetus--Palca;  \citealt{2018ApJ...862..114S}), albeit with a much smaller observed stellar density ($\approx$ 1 degree$^{-2}$). We only detected the disrupted extended structure in Gaia without a clear location of the stream progenitor. Additionally, we detected a second peak at $\phi_2 =-20$~deg near a new stream recently detected in the LAMOST survey, which we later show is separate from Cocytos despite their proximity on the sky. The stream matches the track by  \cite{2009ApJ...693.1118G}, but a first look suggests a larger width than the \cite{2009ApJ...693.1118G} estimate of 0.7$^\circ$. We estimate the stream's width using confirmed spectroscopic members in Section \ref{sec:massMvmet}.

\subsection{Magellan/MagE Follow-up}

We observed a subset of the Cocytos candidate giant stars from \textit{Gaia} with the Magellan Echellette Spectrograph (MagE; \citealt{Marshall2008}) on the 6.5 meter Magellan Baade Telescope at Las Campanas Observatory.  We selected and observed \CocNMageObsRnd{} giants based on the CMD and PM selections shown in Figure~\ref{fig:gaiaselection}, focusing the spectroscopic follow-up on stars with a broad $|\phi_2| < 10^\circ$ requirement, using magnitude-dependent exposure times ranging from $10-20$~minutes per star to achieve signal-to-noise ratio (SNR)~$\approx 10~\mathrm{pixel}^{-1}$ at $5100$~\AA. 

The MagE spectra were reduced and analyzed following the procedures described in several previous works using the same pipelines \citep{2023ApJ...951...26C, Chandra2024}. 
Briefly, spectra were reduced with a fully automated pipeline\footnote{\url{https://github.com/vedantchandra/merlin/releases/tag/v0.2}} built around the \texttt{PypeIt} utility \citep{pypeit:joss_pub}. 
We estimated stellar parameters with the Bayesian \minesweeper~code \citep{Cargile2020} by fitting a region of the MagE spectra (from 4800--5500~\AA) that contains the \ion{Mg}{1} triplet, where our line lists are best-calibrated. 
We also included archival optical--infrared broadband photometry and the Gaia parallax in the likelihood. \minesweeper ~compares these observables to synthetic Kurucz model spectra and additionally constrains the models to lie on \texttt{MIST} isochrones \citep{Kurucz1970,Kurucz1981,Choi2016}. The posterior distribution of stellar parameters is sampled with \texttt{dynesty} \citep{Speagle2020}, producing measurements of the radial velocity $v_\mathrm{r}$, effective temperature $T_\mathrm{eff}$, surface gravity $\log{g}$, metallicity [Fe/H], [$\alpha$/Fe] abundance, and heliocentric distance.  Repeat observations of bright radial velocity standard stars --- HIP4148 and HIP22787 --- over multiple nights indicate that our systematic RV accuracy floor is $\approx 1$\ km s$^{-1}$.

\subsection{RRLs and BHB Members}
We searched for additional candidate members in the Gaia \texttt{vari\_rrlyrae} catalog \citep{2018A&A...618A..30H} using the proper motion and sky selections described in Section \ref{sec:gaiaselection}. To determine their distances, we used the absolute magnitude--metallicity relation by \cite{2018MNRAS.481.1195M} to obtain their absolute $G$ magnitudes and metallicities reported in the Gaia catalog as  
\begin{equation}
    {\rm M}_{\rm G}= 0.32 ~(\pm 0.04) \times {\rm [Fe/H] } + 1.11 ~(\pm 0.06).
\end{equation} 
Uncertainties were determined by propagating the metallicity errors and intrinsic scatter in the relations by Monte Carlo sampling. By selecting RRLs with $|\phi_2| < 5$$^\circ$ and $|\phi_1| < 50$$^\circ$ and requiring distances between 10 and 30 kpc, we obtained two sources with a median distance of \CocRrlDistRnd~kpc. These two sources have not been observed by DESI, but a full radial velocity catalog of DESI RRLs is provided by \cite{2025arXiv250402924M}. We searched for additional members in the DESI BHB catalog \citep{2024arXiv241009149B} using our proper motion selection criteria, requiring $|\phi_2|<5$~deg and $|\phi_1|<50$~deg. This search yielded two BHBs, but their radial velocities were $\approx$-242~km s$^{-1}$ and $\approx$10~km s$^{-1}$, respectively; hence, we do not consider these objects to be members of the stream. 

\section{Stream Characterization}\label{sec:methods}

\subsection{Spectroscopic Members and Galactic Orbit }

\begin{figure*}
    \centering
    \includegraphics[width=\textwidth]{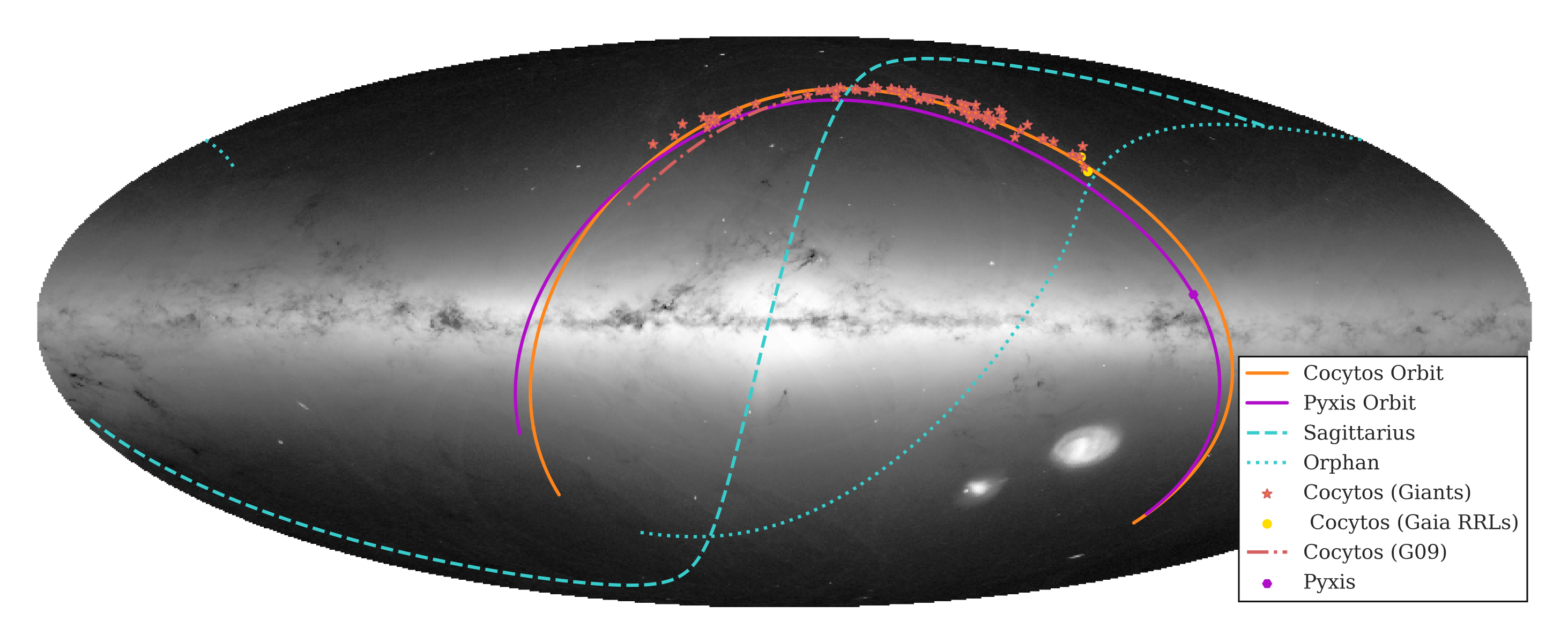}
    \includegraphics[width=\textwidth]{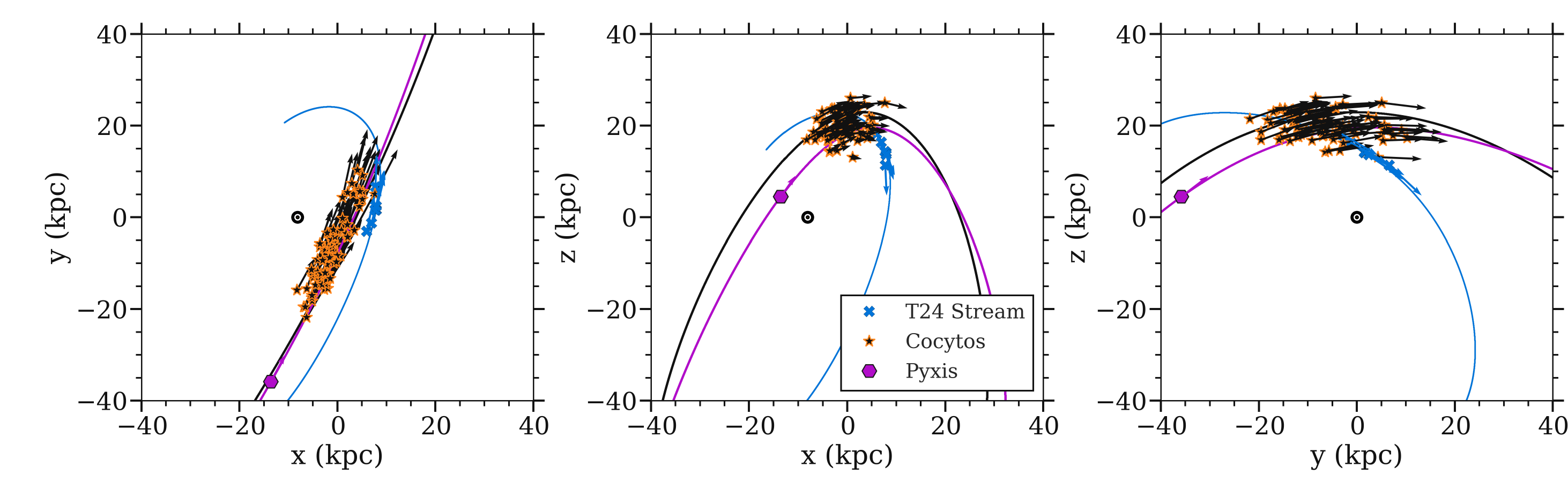}
    \caption{(a) Cocytos orbit (orange) compared to the track by \cite{2009ApJ...693.1118G} (red), tracks of Sagittarius (dashed line) and Orphan (dotted line), and the orbit of the Pyxis cluster (purple line). The background distribution shows binned maps of the Gaia stellar densities. Orange stars are confirmed spectroscopic members with DESI and MagE, and Gaia RRL members are yellow circles.
    (b) 3D-Galactic track of Cocytos (black track and orange-edged stars) compared to T24 \citep{2024ApJ...965...10T} (blue track); crosses show T24 stream members with radial velocity measurements from LAMOST; and the Pyxis globular cluster (purple track and hexagon). Arrows show the velocity direction, and the Sun's position is shown. Cocytos and T24 are clearly distinct structures based on their motions and 3D positions, but Cocytos and Pyxis are on similar orbits.
    }
    \label{fig:orbitparams}
\end{figure*}

\begin{figure*}
    \centering
    \includegraphics[width=\textwidth]{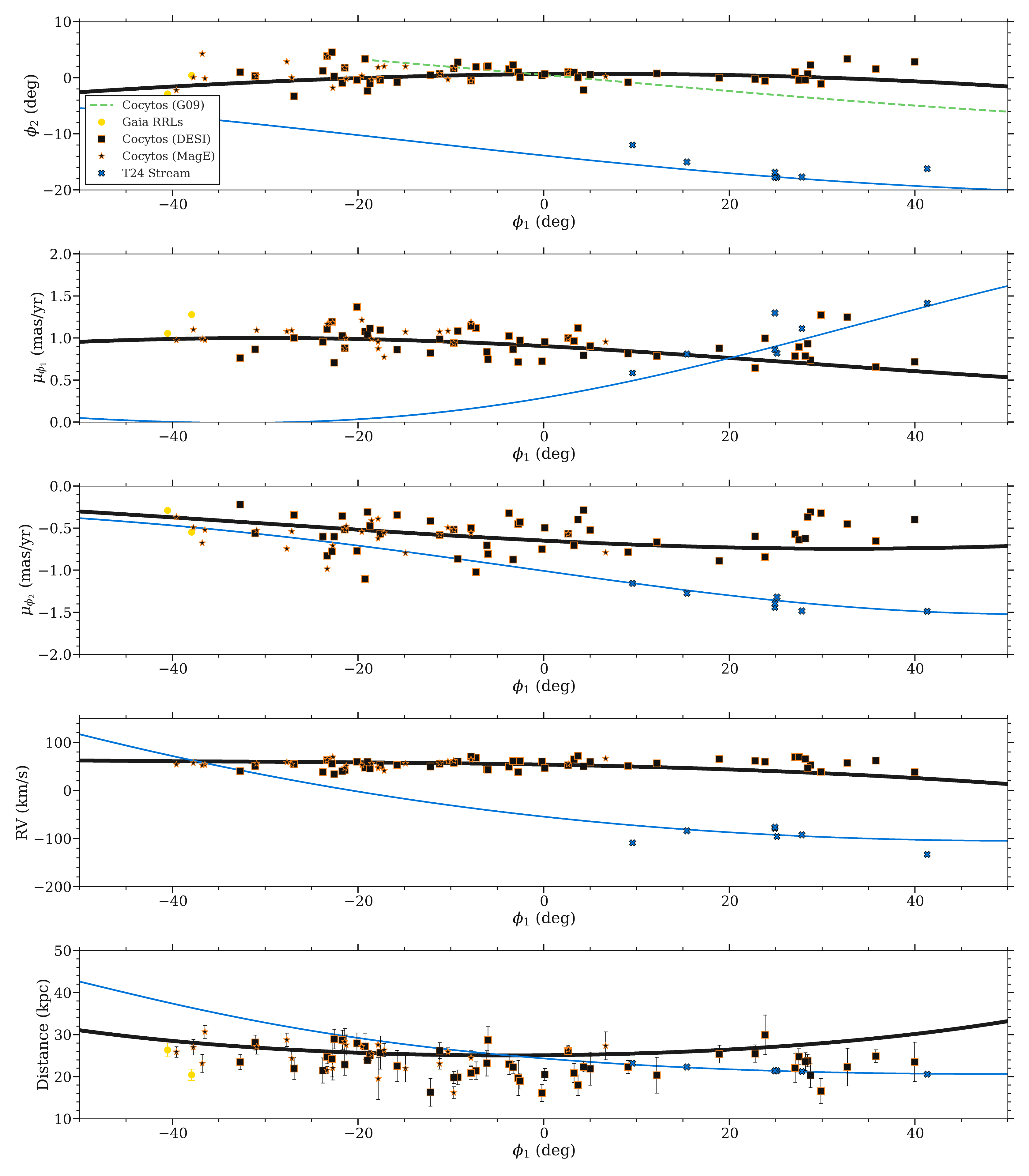}
    \caption{ The stellar orbit of the stream compared to the track by \cite{2009ApJ...693.1118G} and the orbit of the T24 stream recently discovered by \cite{2024ApJ...965...10T} in blue, and T24 members with LAMOST radial velocities are shown as blue crosses. Black rectangles represent spectroscopic DESI members, and MagE members are shown with orange-edged stars. Additionally, we show high-probability Gaia RRL members with yellow circles. }
    \label{fig:phi1track}
\end{figure*}
To compile a final list of spectroscopically-confirmed members, we looked for additional members in DESI year 3 data and combined this sample with our MagE targets by applying the selections described in Section \ref{sec:gaiaselection}. Our membership selection also required $ |\phi_2 | < 5$~deg,  $ |\phi_1 | < 40$~deg, distances between 10$-$30~kpc, and our proper motion selection (see Figure \ref{fig:gaiaselection}), radial velocities between $40-80$~km s$^{-1}$. To select high-confidence members from both MagE and DESI data, we also required a radial velocity uncertainty of $<5$~km s$^{-1}$ and a distance uncertainty $<5$~kpc, and, to remove disk contamination, $V_\phi > -100$ km s$^{-1}$, excluding one additional star. We note that previous studies \citep{2023arXiv231116960S,2025arXiv250213236T} have devised a more robust data-driven stream selection method. However, since our sample size is small, we limited our analysis to this simple, visually informed selection. With these constraints, we obtained a total of \CocNUniqueRnd{} unique members (\CocNDesiRnd{} members in DESI and \CocNMageRnd{} members in MagE).

We computed the stellar orbit from the inverse-variance-weighted mean phase-space coordinates of the combined DESI and MagE giant members. These coordinates yield positions of ($\alpha$, $\delta$, d)=(\CocOrbitRARnd~deg, \CocOrbitDecRnd~deg, \CocOrbitDistRnd~kpc) and velocities ($\mu_{\alpha}, \mu_{\delta*}$, RV)= (\CocOrbitPMRARnd~$\mathrm{mas~yr}^{-1}$, \CocOrbitPMDecRnd~$\mathrm{mas~yr}^{-1}$, \CocOrbitRVRnd~km s$^{-1}$), and Galactocentric positions of (x, y, z)= (\CocXRnd, \CocYRnd, \CocZRnd)~kpc and velocities of (v$_x$, v$_y$, v$_z$)~= (\CocVxRnd, \CocVyRnd, \CocVzRnd)~km s$^{-1}$ \footnote{Throughout this paper, we assumed the R$_{\sun}$=8.122 kpc, Z$_{\sun}$=20.8 pc \citep{2019A&A...625L..10G,2019MNRAS.482.1417B}, with Solar Galactocentric velocities of (V$_x$, V$_y$, V$_z$)=(12.9, 245.6, 7.78)~km s$^{-1}$  ({\tt Astropy v4.0} defaults) }. We integrated these 6D-coordinates backward and forward from -0.5~Gyr to 0.5~Gyr using the \texttt{MilkyWayPotential2022} Galactic potential with \texttt{Gala}. We obtained an orbital apocenter of \CocApoRnd~kpc, a pericenter of \CocPeriRnd~kpc and an eccentricity of \CocEccRnd from the resulting orbit. We list orbital parameters and other stream properties in Table \ref{tab:params}.  Figure \ref{fig:orbitparams} shows the orbit of Cocytos compared to other nearby streams, and the original track by \cite{2009ApJ...693.1118G}. Our computed orbit overlaps and is nearly parallel to the track by \cite{2009ApJ...693.1118G}; however, their orbit does not fully cover the extent of the entire stream ($\phi_1< 0$~deg). Additionally, our estimate of the heliocentric distances of $\approx$~\CocOrbitDistRnd~kpc is discrepant with their isochrone-based estimate of $\approx$~11~kpc.  

Additionally, Cocytos overlaps with other known streams on the sky, namely, Sagittarius and Orphan. Recently, \cite{2024ApJ...965...10T} detected a stream (hereafter, the T24 stream) at a similar sky location and with proper motions similar to Cocytos, but a negative radial velocity of $\approx-87$~km s$^{-1}$ and a mean metallicity [M/H]$=-1.3$. We compare this stream's orbit to Cocytos, which runs parallel to the T24 stream. Interestingly, hints of this stream could be visible in our Gaia selection as a secondary peak below Cocytos ($\phi_1\approx $50~deg, $\phi_2\approx $-20~deg), as shown in Figure \ref{fig:gaiaselection}. 

Very recently, \cite{2025arXiv250322979T} announced the discovery of a stream near Sagittarius with proper motion consistent with Cocytos, a length of $\approx$~110 degrees, a width of $\approx$1.2~kpc, a median [Fe/H]$=-1.3$, and at a distance of 27~kpc.  This new stream is consistent with our discovery. However, with the lack of complete information on the radial velocity of its members, they estimate a dispersion of $\approx$22.4 km s$^{-1}$ based on tangential velocities. They do not associate this stream with Cocytos, but they associate it with the globular cluster Pyxis. Pyxis is not on the DESI footprint, but it has a metallicity of $-1.2$, a radial velocity of 40 km s$^{-1}$ and is located at 38~kpc \citep{2021MNRAS.505.5978V}. As shown in Figure \ref{fig:orbitparams}, T24 and Cocytos are separate structures, but Cocytos and Pyxis have very similar orbits.

Figure \ref{fig:phi1track} shows the track, proper motion, and radial velocity along $\phi_1$ compared to the T24 stream, with giants and likely Gaia RRL members shown separately. We separate different datasets for a pipeline-independent analysis. 
DESI giant members have median distance uncertainties of $\approx$~\CocDesiDistErrRnd~kpc, RV  uncertainties of ~\CocDesiRVErrRnd~km s$^{-1}$ (with an additional 1 km s$^{-1}$ systematic) which are higher than MagE distance errors of $\approx$~\CocMageDistErrRnd~kpc and RV uncertainties of \CocMageRVErrRnd~km s$^{-1}$ (with an additional 1 km s$^{-1}$ systematic). Additionally, the T24 track crosses the Cocytos track in proper motion space for $\phi_1 > 0$~deg, but the two streams are distinct in radial velocity space: Cocytos has RV$>$0~km s$^{-1}$, and T24 has RV$<-24$~km s$^{-1}$ for $\phi_1 > 20$~deg. However, these two structures (Cocytos and T24) are intertwined, and visually, Cocytos is thicker. Overall, we notice a large scatter of individual giant members compared to the computed median orbit, which could be interpreted as a signature of kinematic heating. Later in the paper, we formulate the hypothesis that Cocytos is likely a GC stream that fell into the Milky Way with the GSE; it has been shown that GC streams that infell with a dwarf galaxy are kinematically heated \citep{2021MNRAS.501..179M}. A detailed chemical abundance analysis and full kinematic modeling are needed to confirm this hypothesis.

\begin{figure*}
    \centering
    \includegraphics[width=\textwidth]{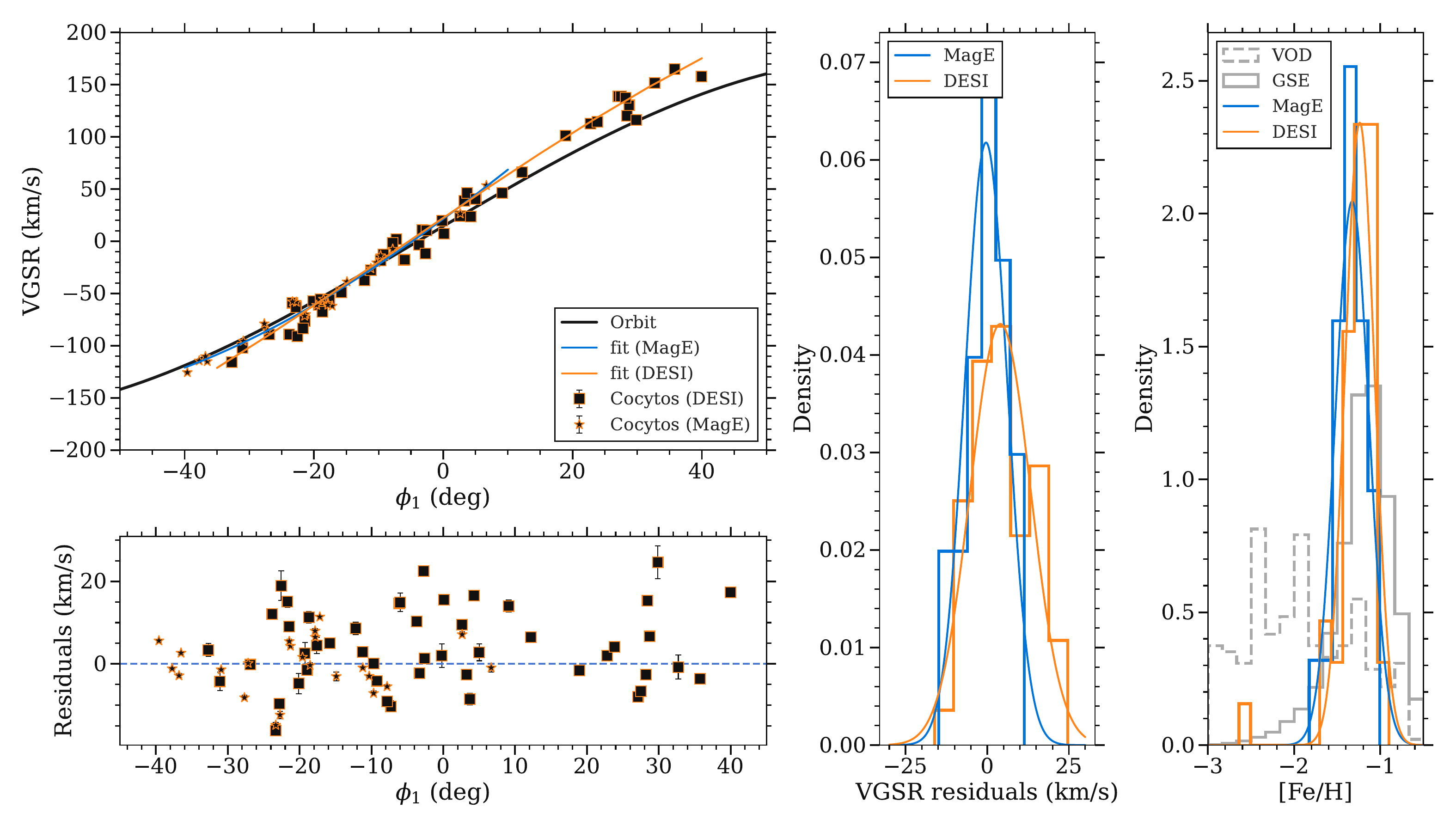}
    \caption{Measurement of the velocity and metallicity dispersions of stream members. {\it Top Left}:  $V_{\rm GSR} $ distribution of members with DESI spectra (squares) and MagE spectra (stars). The black line shows our computed orbit of the stream, and the blue and orange lines show the 3rd-degree polynomial fit to the MagE and DESI data separately. {\it Bottom Left}: Residual differences (fit-data) between the interpolated track of the stream and  $V_{\rm GSR} $  values for DESI members (squares) and MagE members (stars). {\it Center}:  Histograms show the distribution of the $V_{\rm GSR} $ difference between the interpolated track and stream members, while smooth distributions are Gaussians fit to the data by clipping 3-$\sigma$ outliers.
    {\it Right}: Metallicity distribution of Cocytos with MagE spectra (blue) or DESI spectra (orange) compared to GSE (grey solid lines) obtained from (Carrillo et al. in prep) and the VOD (grey dashed lines) obtained from \cite{2016ApJ...831..165V}.  We show the best-fit Gaussian to the metallicity distribution of the stream with DESI and MagE. }
    \label{fig:rv_fits_example}
\end{figure*}

\subsection{ Velocity Dispersion, Metallicity Dispersion, and Stream Width }\label{sec:massMvmet}
\begin{figure*}
    \centering
    \includegraphics[width=\textwidth]{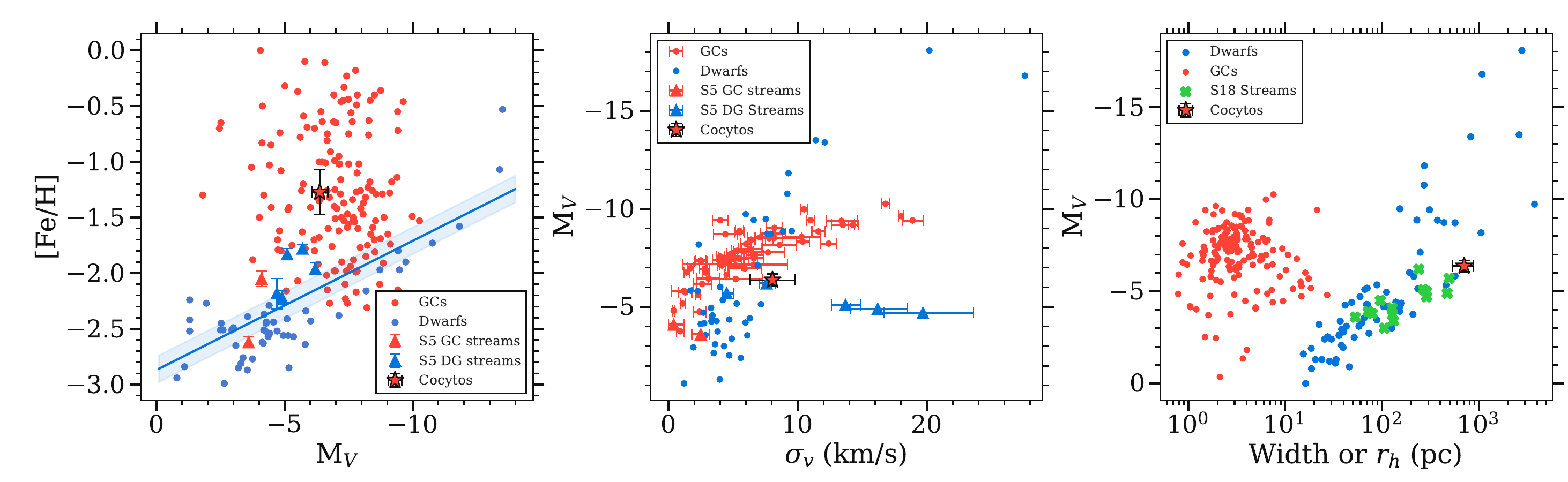}
    \includegraphics[width=\textwidth]{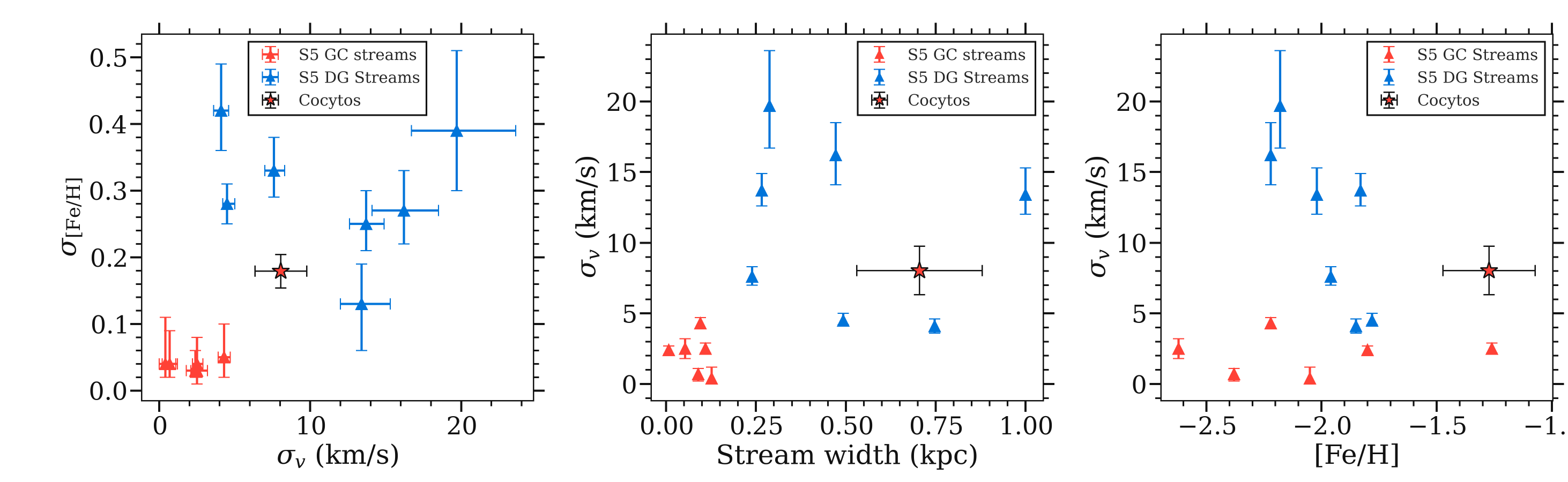}
    \caption{Comparison between Cocytos and globular clusters \citep{2014ApJS..210...10R} shown as red points, Milky Way dwarfs from the Local Volume Database \citep{2024arXiv241107424P} shown as blue points and stellar streams in the S5 survey \citep{2022ApJ...928...30L} shown as triangles and DES streams \citep{2018ApJ...862..114S} are shown as green crosses. In the S5 sample, streams with metallicity dispersion $<0.2$~dex at a 95~\% confidence level, were labeled as likely to have globular cluster progenitors. We show these streams as red triangles and those with DG progenitors as blue triangles. We used the M$_{\rm V}$ values and stream widths reported by \cite{2018ApJ...862..114S} for streams in both the S5 and the DES catalogs and combined $\sigma_v$ and [Fe/H] measurements for Cocytos using DESI and MagE members. {\it Top:} M$_{\rm V}$, [Fe/H], velocity dispersion, and stream widths compared to our measurements for Cocytos shown as a red star. We show the mass-metallicity relation for Milky Way dwarfs by \cite{2013ApJ...779..102K} with 1-sigma contours in blue. Our measurement of [Fe/H]$=\CocFeHRnd$ is consistent with a GC progenitor, as a typical DG stream would have a higher luminosity at this metallicity.
    {\it Bottom:} Comparison between Cocytos and other stellar streams. Our measured velocity dispersion and width are higher than those of other GC streams but the combination of the metallicity dispersion, luminosity, and velocity dispersion suggests it is likely a GC stream.
    }
    \label{fig:compGCsDGs}
\end{figure*}

To estimate the internal velocity dispersion, the spread in the metallicity, and the stream's width, we fit Gaussians to the data, separating MagE and DESI giants to mitigate systematics between different pipelines. To estimate the internal velocity dispersion of the stream, we first estimated the stream track by fitting members in $\phi_1-V_{\rm GSR}$~space with a 3rd-degree polynomial. After this step, we computed the residual differences between the stream members $V_{\rm GSR}$ and the interpolation. Similarly to the method used by \cite{2022ApJ...928...30L}, we fit the residuals to a Gaussian using a Markov chain Monte Carlo fitting routine implemented in the \texttt{EMCEE}  \citep{2013PASP..125..306F} package. We used a set of 50 walkers and 10,000 steps with 1,000 burn-in steps by repeatedly discarding $3\sigma$~outliers with a uniform prior on the central radial velocities $\in$[-50 , 50]~km s$^{-1}$, a log-uniform prior $\log \sigma_v$ $\in$ [-2, $\log$~20]~km s$^{-1}$, and we added a $1$~km s$^{-1}$ systematic wavelength calibration error to the reported random errors of DESI and MagE \citep{2023ApJ...947...37C,2024MNRAS.533.1012K}. With this setup, we obtained $\sigma_v$= $\CocSigmaVMageFmt$~km s$^{-1}$ with MagE and $\sigma_v$=$\CocSigmaVDesiFmt$~km s$^{-1}$ using DESI giants. Concatenating the posterior distribution samples of DESI and MagE, we obtained the $\sigma_v= \CocSigmaVFmt$~km s$^{-1}$. Similarly, we estimated the metallicity and the metallicity spread of the stream by fitting Gaussians to individual metallicity members with a uniform prior (-5 dex to 5 dex) on the metallicity and a log-uniform prior (-2, 0) on the dispersion, with an additional 0.1 ~dex systematic. We obtained [Fe/H]$=\CocFeHDesiFmt$ (with DESI), [Fe/H]$= \CocFeHMageFmt$ ( with MagE)  and $\sigma_{\rm[Fe/H]}=\CocFeHDispDesiFmt$ (with DESI) and $\sigma_{\rm[Fe/H]}=\CocFeHDispMageFmt$ (with MagE). Figure \ref{fig:rv_fits_example} shows our measurements of the velocity and metallicity dispersions of the stream using a Gaussian fit to the $V_{\rm GSR} $ distribution of stream members. We notice that MagE members produce a lower velocity dispersion compared to DESI, and the DESI members have a double-peaked distribution in the velocity residuals, but we do not fit a higher order polynomial to these stars to avoid overfitting. The discrepancies between the median metallicity and velocity dispersions for MagE and DESI targets may be attributed to differences in data quality or spectral reduction and fitting pipelines. Nevertheless, the measured metallicity dispersions are consistent between DESI and MagE giants within their standard deviations. Finally, we estimated the size of the stream by separately fitting a Gaussian to the $\phi_2$ distribution of DESI and MagE Cocytos members. We obtained a combined Gaussian width of $\phi_2$ of $\CocWidthDegFmt$~$^\circ$, which corresponds to a physical width of $\CocWidthKpcFmt$~kpc at a distance of $\CocOrbitDistRnd \pm 5$~kpc.  

To compare these measurements to other streams, we obtained velocity dispersion measurements and stellar metallicities for thirteen spectroscopically-followed-up streams by the S5 collaboration (\citealt{2022ApJ...928...30L}: 300S, AAU, Chenab, Elqui, Indus, Jet, Jhelum, Ophiuchus, Orphan, Phoenix, Sagittarius, Turranburra, WillkaYaku). Additionally, we obtained velocity dispersion, metallicities, and metallicity dispersion measurements for globular clusters in the \cite{2014ApJS..210...10R} catalog and a list of Milky Way dwarf galaxies in the Local Volume Database \citep{2024arXiv241107424P}. Figure \ref{fig:compGCsDGs} compares our values of $\sigma_v$, ${\rm [Fe/H]}$ and $\sigma_{\rm [Fe/H]}$ to S5 streams that are classified as having globular cluster or dwarf galaxy progenitors. Globular cluster streams in S5 typically have $\sigma_v < 5$~km s$^{-1}$ and $\sigma_{\rm [Fe/H]} < 0.1 $ dex. Although our measured dispersions are higher than the cut-off point from the S5 survey, its high overall metallicity hints at a globular cluster progenitor. Recently, \cite{2025ApJ...980...71V} measured a 5--8 km s$^{-1}$ cocoon feature (a higher velocity dispersion distribution of stars) around cold ($\approx$3~km s$^{-1}$) GD-1 with DESI data. We do not see a clear signature of such a feature in Cocytos. Additionally, we do not detect a clear location of the progenitor. Nevertheless, we claim Cocytos likely had a globular cluster progenitor by comparing the metallicity, the stellar mass and luminosity of the stream to known GCs and dwarfs in Section \ref{sec:mass}.

\subsection{Stellar mass and luminosity}\label{sec:mass}

We used the sample of proper-motion-selected Gaia stars described in Section \ref{sec:gaiaselection} to estimate the stellar mass, absolute magnitude and stream width. We selected a Kroupa-IMF-sampled isochrone from the PARSEC grid \citep{2012MNRAS.427..127B}.  Following the method by \cite{2018ApJ...862..114S} and other previous studies, we estimated the M$_{\rm V}$ of the stream by summing up the V-band fluxes of each Gaia-selected member, corrected by the unobserved luminosity of the stream beyond our magnitude limit, which is estimated based on an isochrone. We converted its observed Gaia G magnitudes and BP$-$RP colors into V-band magnitudes as
\begin{equation}
G-V= -0.0176-{\rm (BP-RP)}\times 0.006-0.1732\times {\rm (BP-RP)} ^2
\end{equation} with an intrinsic scatter of 0.04.\footnote{Conversions provided by the Gaia data releases: \url{https://gea.esac.esa.int/archive/documentation/GDR2/Data_processing/chap_cu5pho/sec_cu5pho_calibr/ssec_cu5pho_PhotTransf.html}}
To compute the unobserved luminosity of the stream, we summed up the mass-weighted M$_{\rm V}$ of the isochrone for all stellar members in Gaia with G$<19$~mag, assuming 68 percent (1-sigma) of the total luminosity is contained within the stream width. We computed the total M$_{\rm V} $ by correcting the observed luminosity by the unobserved luminosity offset to the distance modulus of the stream. We also estimated the mass of the stream by summing up the individual masses in the IMF-sampled isochrone normalized to the number of observed stars for Gaia G mag$<19$. To estimate the uncertainty in the mass and $M_{\rm V}$ of the stream, we propagated errors in the BP$-$RP colors, stream width, metallicity, age, distance, and G mags by Monte Carlo sampling. We did not use a single value for the $\rm {[Fe/H]}$, age, and distance of the stream to select the stream isochrone. Instead, we used a uniform distribution of metallicities between $-1$ and $-2$, a uniform distribution of ages between 10~Gyr and 13~Gyr, and a uniform distribution of distances between 15~kpc and 30~kpc, which reasonably match our Gaia selection. While these ranges are broad, they provide a conservative estimate of the structural properties of the stream. This procedure results in $M_{*}=\CocMstarFmt \times 10^4$ \Msun and $M_{\rm V}= \CocMvFmt$. Using a more stringent set of assumptions by drawing distances and metallicities as a normal distribution centered around the measured values of the median and standard deviation of MagE and DESI giant members but keeping the age distribution as uniform between 10~Gyr and 13~Gyr, we obtained $M_{*}=\CocMstarRelFmt\times 10^4$ \Msun and $M_{\rm V}= \CocMvRelFmt$. As we did not select a large number of stars at the main-sequence turn-off of the stream, we could not independently obtain a best-fit isochrone for the stream or an estimate of the cluster's age. These results provide a lower limit on the luminosity of the stream as it could be extended in $\phi_1$ beyond our cutoff of 40~$^\circ$. The high mass and luminosity of the stream are consistent with its earlier detection in SDSS.

To contextualize these measurements, we compare our estimated mass, velocity dispersion, [Fe/H], and $M_{\rm V}$ of the stream with a set of known Milky Way globular clusters and dwarf galaxies and their streams. Figure \ref{fig:compGCsDGs} shows this comparison. Starting with the M$_{\rm V}$ and $\rm{[Fe/H]}$ of the stream, we find that these values are more similar to Milky Way globular clusters than dwarf galaxies, as the high $\rm{[Fe/H]}$ of Cocytos requires a substantially larger mass and luminosity than a dwarf galaxy progenitor. Our estimated luminosity puts it at $\approx 2\sigma$ above the [Fe/H]-luminosity relation for Milky Way dwarfs by \cite{2013ApJ...779..102K} (see Figure \ref{fig:compGCsDGs}). However, for Cocytos to be a remnant of a dwarf galaxy, the mass-metallicity relation requires a stellar mass of $\approx 2 \times 10^7$~\Msun, making it at least a tenth of the mass of Sagittarius \citep{2005ApJ...619..807L}. On the other hand, the M$_V=\CocMvFmt$ makes its progenitor one of the brightest globular clusters and one of the most metal-rich known globular cluster streams in the Galaxy. It is possible that the metallicity dispersion may still be overestimated, given our metallicity precision of $\approx$0.1~dex. Additionally, Cocytos has one of the largest apocenters among the streams discovered to date \citep{2024arXiv240519410B}. Comparing the stream's width to DES streams, we find that Cocytos is thicker than other streams identified in the S5 survey and the thickest out of all the globular cluster streams in S5. This thickness could be tentatively attributed to pre-heating in its parent dwarf before infall \citep{2021MNRAS.501..179M} or to an imperfect membership selection methodology. Although the dwarf galaxy progenitor hypothesis could explain the thickness, the high metallicity requires an extremely massive stream, making this scenario more unlikely. Future work will determine the metallicity and velocity dispersion to a higher precision with higher-resolution spectroscopy.

\begin{deluxetable*}{lcc}
\tabletypesize{\small}
\tablecaption{Summary of the structural properties of the Cocytos stream \label{tab:params}}
\tablehead{
\colhead{Quantity} &
\colhead{Value} &
\colhead{Unit}}
\startdata
R.A. & \CocRARnd (150, 250)\tablenotemark{a} & $\mathrm{deg}$ \\
Dec & \CocDecRnd (-40, 20)\tablenotemark{a} & $\mathrm{deg}$ \\
$\mu_{\alpha}$ & $\CocPMRAFmt$ & $\mathrm{mas~yr}^{-1}$ \\
$\mu_{\delta }*$ & $\CocPMDecFmt$ & $\mathrm{mas~yr}^{-1}$ \\
Radial velocity & $\CocRVFmt$  \tablenotemark{b}  & $\mathrm{km~s}^{-1}$ \\
Heliocentric Distance  & $\CocDistFmt$ & $\mathrm{kpc}$ \\
Apocenter & $\approx \CocApoRnd $ & $\mathrm{kpc}$ \\
Pericenter & $ \approx \CocPeriRnd $ & $\mathrm{kpc}$ \\
Eccentricity & $\approx \CocEccRnd $ & \\
$L_z$ & $\CocLzFmt$ & $10^3~\mathrm{km~kpc~s}^{-1}$ \\
$E_{\mathrm{tot}}$ & $\CocEtotFmt$ & $10^5~\mathrm{km}^2~\mathrm{s}^{-2}$\\
\hline
$M_{\mathrm{V}}$ & $\CocMvFmt$ \tablenotemark{c} & \\
$M_*$ & $\CocMstarFmt$ \tablenotemark{c} & $\times 10^4~\mathrm{M}_\odot$ \\
\rm{[Fe/H]} & $\CocFeHMageFmt$ (MagE) & \\
 & $\CocFeHDesiFmt$ (DESI) & \\
 & $\CocFeHTableFmt$ (combined) & \\
$\sigma_{\mathrm{[Fe/H]}}$ & $\CocFeHDispMageFmt$ (MagE) & \\
 & $\CocFeHDispDesiFmt$ (DESI) & \\
 & $\CocFeHDispFmt$ (combined) & \\
$\sigma_v$ & $\CocSigmaVMageFmt$ (MagE) & $\mathrm{km~s}^{-1}$ \\
 & $\CocSigmaVDesiFmt$ (DESI) & $\mathrm{km~s}^{-1}$ \\
 & $\CocSigmaVFmt$ (combined) & $\mathrm{km~s}^{-1}$ \\
Width & $\CocWidthKpcFmt$ & $\mathrm{kpc}$ \\
\enddata
\tablecomments{Quantities in this table were computed as the median and standard deviation of confirmed spectroscopic members, except orbital apo/peri/ecc and Gaussian-fit quantities, which are described in the text and footnotes.}
\tablenotetext{a}{Quantities in parentheses show the full range of the stream selected with Gaia proper motion and isochrone filter as described in Section~\ref{sec:gaiaselection}.}
\tablenotetext{b}{These are the median and standard deviation for the radial velocities of confirmed spectroscopic members. We separately compute the internal velocity dispersion by fitting a Gaussian and masking outliers (sigma-clipping) as described in Section \ref{sec:massMvmet}.}
\tablenotetext{c}{Quantities were estimated using Gaia-selected members.}
\end{deluxetable*}

\begin{figure*}
    \centering
    \includegraphics[width=\textwidth]{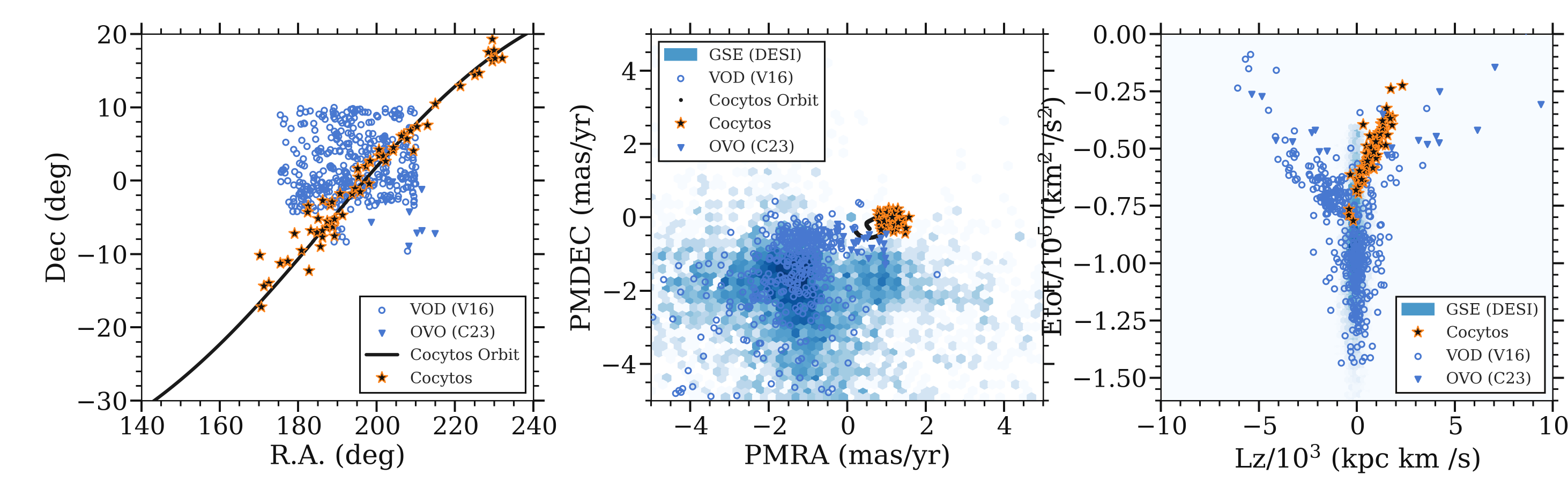}
    \includegraphics[width=\textwidth]{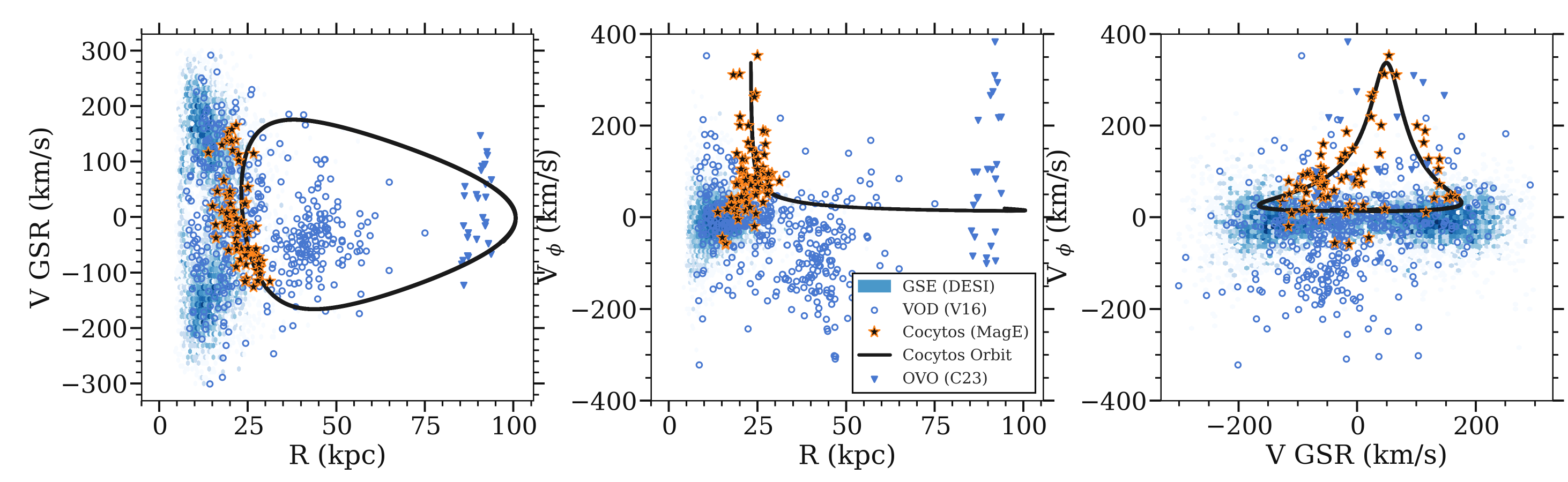}
    \caption{Comparing Cocytos to the VOD and GSE. Black stars with orange contours show our final Cocytos members, blue circles are RRL members of the VOD by \cite{2016ApJ...831..165V}, blue hexbin shows kinematically selected GSE giant members in DESI by Carrillo et al. in prep, and the black line shows the orbit of Cocytos. {\it Top Left:} Sky positions showing the overlap between Cocytos and the VOD {\it Top Center:} Proper motion distribution. We observe strong clustering in proper motion for Cocytos members compared to the broader VOD and GSE distribution. 
    {\it Top Right:} The distribution in L$_z$-E space shows a significant overlap between VOD and GSE with a few outliers in VOD that overlap with our final Cocytos members. This supports the idea that VODs are likely outer shells of GSE. {\it Bottom:} Distribution of Galacto-centric radial distance and Galacto-centric velocities. There is significant overlap between the VOD and GSE.
    }
    \label{fig:comptoGSE}
\end{figure*}

\section{Discussion}\label{sec:discuss}

In this section, we discuss Cocytos in relation to the other streams and substructures detected to date. Cocytos is located near the Virgo Overdensities (VOD) on the sky. The VOD were named by \cite{2008ApJ...673..864J}, who identified an extended clump of stars in the Virgo constellation in SDSS and 2MASS covering over 1,000 sq. deg at distances between 6--20~kpc. There are other sub-structures in this region of the sky, namely, the Virgo Stellar Stream (VSS, \citealt{2006ApJ...636L..97D}),  the Outer Virgo Overdensities (OVO, \citealt{2017ApJ...844L...4S}), the Parallel stream \citep{2016ApJ...833..235S} and the Perpendicular stream \citep{2019MNRAS.486.2618B}, and possibly-connected structures in other parts of the sky with unknown origins, namely, the Hercules--Aquila Cloud (HAC, \citealt{2007ApJ...657L..89B}) and the Eridanus--Phoenix (EriPhe) overdensities \citep{2016ApJ...817..135L}.

Now, we turn to discussing interpretations of these structures. The most dominant interpretation of the VOD is that they are debris of a past merger \citep{2008ApJ...673..864J,  2012AJ....143..105B}. \cite{2016ApJ...817..135L} proposed that VOD, HAC, and EriPhe might have originated from apocenter pile-ups of a massive radial merger given they are approximately at similar Galactocentric distances 120$^\circ$ apart on one orbital plane before GSE was discovered. 
\cite{2019MNRAS.482..921S} suggested a common origin of the VOD and HAC, given their substantial overlap in orbital parameters/integrals of motion space. With the identification of these disparate groups and streams at the same region in the sky and distance, \cite{2019ApJ...886...76D} proposed that they could all be explained by one $10^9$ \Msun~dwarf galaxy merger and a separate stream with a track that is consistent with Cocytos using tailored N-body simulations. They named this merger the ``Virgo Radial Merger." The debris of the Virgo radial merger
extends from the solar neighborhood to distances of $\sim$60 kpc.
\cite{2021ApJ...923...92N} extensively explored the formation of GSE using 500 tailored N-body simulations constrained on H3 data. They reproduced the spatial properties of the VOD and the HAC as apocenter pile-up of GSE debris, which also resulted in a doubly broken power-law profile for the stellar halo \citep{2022AJ....164..249H}. This picture is further confirmed by precise measurements of the velocities of a large sample of giant stars in VOD and OVO by \cite{2023ApJ...951...26C}. However, \cite{2020ApJ...902..119D} argue that shells of GSE would dissipate in $\sim$5 Gyr; hence, based on this timing argument, it is necessary to invoke a different merger event separate from GSE to explain the VOD \& HAC structures.  \cite{2021AA...654A..15B} integrated the orbits of local ($<$2.5~kpc) halo stars that are associated with confirmed other prominent halo substructures generally associated with mergers (GSE, Sequoia, Helmi streams, and Thamnos) and found that overdensities in the VOD and HAC regions are consistent with a pile-up of orbits in these structures. However, these previous results used a spherical potential. \cite{2022ApJ...934...14H} showed that the VOD and HAC persist for billions of years by integrating their orbits in a tilted potential. Studies with a suite of a large number of cosmological simulations with a GSE-like merger (e.g., the MWest suite, \citealt{2024arXiv240408043B}) would further help elucidate the nature of stellar overdensities in the halo and their connection to mergers.  

We suggest that Cocytos is related to VOD and the GSE based on three main points: co-location on the sky and distance, a radial and eccentric orbit with orbital energies similar to GSE and its associated GCs, and a similar median metallicity with GSE and the VOD. To compare the Cocytos stream to GSE and the VOD, we obtained a sample of kinematically-selected GSE giant stars in DESI year 1 data (Carrillo et al. in prep.), and the full 6D positions of VOD RR Lyrae by \cite{2016ApJ...831..165V} and giant members of the  OVO (\citealt{2023ApJ...951...26C}). As the OVO are located $>50$~kpc, the distance uncertainties are likely significant, so we assumed a fixed distance of $\approx$90~kpc. We computed orbital properties for these structures with the \texttt{MilkyWayPotential2022} Galactic potential implemented in \texttt{Gala}. Figure \ref{fig:comptoGSE} compares the VOD, GSE, and Cocytos. Cocytos overlaps on the sky with the VOD, but it is a separate stream-like feature with a much tighter clustering in proper motion space. The 3D positions and kinematics reveal that most of the VOD is consistent with being shells of GSE. Moreover, while the OVO is located at a greater distance, their radial and azimuthal velocities overlap with Cocytos' orbit and GSE suggesting that the VOD is not separate from the GSE and that the Cocytos stream may be kinematically associated with both the VOD and GSE. Finally, Cocytos has the same median metallicity as GSE, further strengthening this relationship. A more targeted dynamical model and detailed chemical abundance analysis of the VOD, OVO and Cocytos may further elucidate their connection. As shown in figures \ref{fig:comptoGSE} and \ref{fig:contamination}, the VOD, parallel and perpendicular streams are located near Cocytos on the sky, and the OVO overlaps with Cocytos orbit. 

\begin{figure*}
    \centering
     \includegraphics[width=\textwidth]{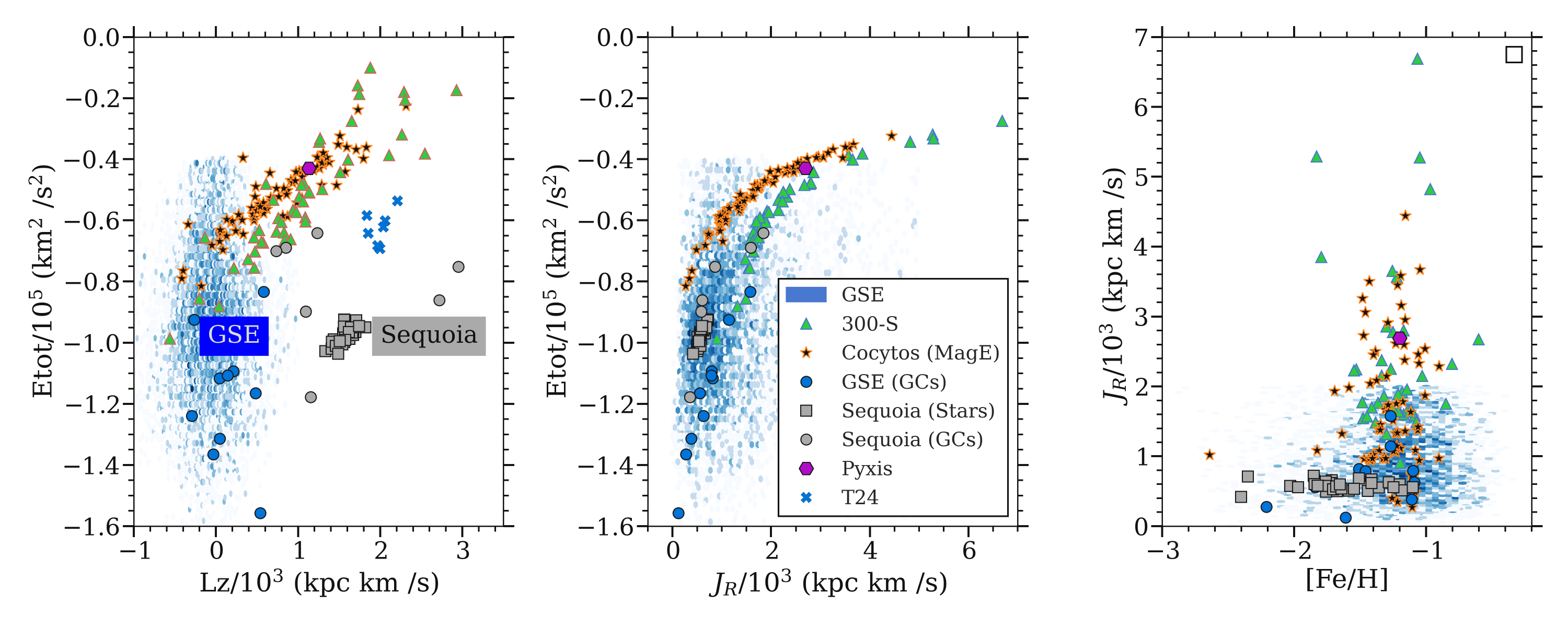}
    \caption{Comparison between globular clusters associated with GSE (blue circles, \citealt{2022ApJ...935..109L}), the 300-S members obtained by the S5 survey (green triangles, \citealt{2022ApJ...928...30L}), DESI GSE members (blue, Carrillo et al. in prep.), the T24 stream (blue crosses, \citealt{2024ApJ...965...10T}), and the Pyxis globular cluster (purple hexagon). We also show DESI data for Sequoia stars (grey rectangles, Kim et al. in prep.) and globular clusters associated with Sequoia (grey circles, association by \citealt{2019A&A...630L...4M}, data taken from the Baumgardt catalog). Comparison in Energy-$L_z$ space ({\it left}), Energy-$J_R$ space ({\it center}), and $J_R$-metallicity space ({\it right}). Cocytos and 300-S have higher total energies than GSE debris, but Cocytos, 300-S and GSE GCs have similar $J_R$ and $L_z$ values, pointing to likely the same progenitor. The T24 stream may have a different origin despite being located near the same galactocentric distance and having a similar metallicity to Cocytos. As Cocytos has a higher $J_R$, $E_{\rm tot}$, and is narrower in metallicity compared to Sequoia, its association with GSE is more likely than its association with Sequoia. The Cocytos stream is also associated with Pyxis globular cluster. }
    \label{fig:comparison_GSE_Jr}
\end{figure*}

We further tie Cocytos to GSE by comparing it to other streams and GCs associated with GSE in the literature.  With a stellar mass of $\approx 10^8$\Msun \citep{2018Natur.563...85H,2020ApJ...901...48N}, GSE is expected to host 5--20 globular clusters with a total stellar mass of $\approx 10^6$\Msun~\citep{2023MNRAS.522.5638C}. Our estimate of the stellar mass for Cocytos would not be inconsistent with this model.
\cite{2022ApJ...935..109L} associated several globular clusters with GSE based on their orbits: NGC 288, 362, 1851, 1904, 2298, 6229, 6341, 7089 and Omega Cen. To compare to another major accreted substructure, namely, Sequoia, we obtained full 6D information of Sequoia-selected giant stars (Kim et al. 2024) and a list of its most-likely associated globular clusters: NGC 3201, NGC 5466, NGC 6101, NGC 7006, Pal 13, IC 4499 and FSR 1758 \citep{2019A&A...630L...4M} with position and kinematic information compiled by \cite{2018MNRAS.478.1520B,2021MNRAS.505.5978V}\footnote{\url{https://people.smp.uq.edu.au/HolgerBaumgardt/globular/orbits.html}}. As shown in Figure \ref{fig:comparison_GSE_Jr}, Sequoia extends to the metal-poor regime and has overall lower energies and a different radial action ($J_R$) compared to Cocytos. Hence, these structures are unlikely to be associated. However, we note that associating globular clusters or streams to accreted structures that built up the Milky Way halo is not yet an exact science, and more work is needed to standardize this process \citep{2024OJAp....7E..23C}. Therefore, the high $({\rm [Fe/H]}=\CocFeHRnd)$, fairly narrow metallicity distribution ($\CocFeHDispRnd$~dex) of Cocytos and its overlap in energy and L$_z$ with GSE makes this association more likely than an association with Sequoia, which extends to much lower metallicity and has a different $J_R$ distribution. As GSE is a massive dwarf galaxy, it would sink closer to the Galactic center due to dynamical friction, and its globular clusters would be ejected or orbit in outer shells of the structure. This picture is corroborated by the overlap with the OVOs (shells of GSE), but a more detailed and tailored N-body model would be needed to reproduce this picture.
 
Interestingly, while Cocytos ([Fe/H]$=\CocFeHRnd$) is more metal-rich than Omega Cen ([Fe/H]$=-1.6$), we measured a similarly high $\sigma_{\rm [Fe/H]}$ of $\CocFeHDispDesiFmt$ dex.  It has been suggested that this high metallicity dispersion for Omega Cen might point to it being a stripped nuclear star cluster \citep{2003MNRAS.346L..11B,2019A&A...630L...4M,2021MNRAS.500.2514P,2020A&ARv..28....4N}.  The 300-S stream has also been proposed to be associated with GSE \citep{2009ApJ...692.1464G,2011ApJ...733...46S,2022ApJ...928...30L}. Its orbit is highly eccentric, which indicates a likely accreted nature \citep{2018ApJ...866...42F,2022ApJ...928...30L}, and it has a metallicity similar to GSE.
\cite{2022ApJ...928...30L} measured [Fe/H]$=-1.26$, with $\sigma_{\rm [Fe/H]}={0.04}_{-0.02}^{+0.04}$ dex, with a small internal $\sigma_v$=${2.5}\pm {0.4}$ km s$^{-1}$, and a width of 110~pc using $\approx$ 53 members. We obtained the positions and kinematics of 300-S members \citep{2022ApJ...928...30L}, GSE globular clusters data from \cite{2022ApJ...935..109L}, and data for the nearby T24 stream obtained from Gaia and LAMOST \citep{2024ApJ...965...10T}. Figure \ref{fig:comparison_GSE_Jr} compares 300-S, GSE, GSE GCs, and the T24 stream. Cocytos and 300-S have the same Energy and $L_Z$ values, and they both significantly overlap with globular clusters associated with GSE in $J_R-L_z$ space, but slightly different radial actions, pointing to possibly the same accretion history. However, the T24 stream has a much higher $|L_z|$ than Cocytos and 300-S. Hence, this stream probably has a different origin despite its proximity on the sky to Cocytos. While Cocytos has a similar metallicity as 300-S, it has a much larger width and our analysis points to a relatively high metallicity dispersion. Recently, \cite{2024MNRAS.529.2413U} used Magellan/MIKE high-resolution spectroscopic follow-up of the 300-S stream to measure evidence of light-element abundance variation, which is usually used to confirm globular cluster progenitors \citep{2018ARA&A..56...83B} and more precisely measure its metallicity dispersion. \cite{2024MNRAS.529.2413U} did not detect a significant metallicity dispersion in 300-S, but they identified a star with enhanced Na abundance, confirming that its progenitor is likely a globular cluster. We further compare the orbital parameters of Pyxis with GSE and Sequoia. We find that Cocytos and Pyxis also overlap in Energy, $L_z$ and $J_R$ further confirming their association. It has been suggested that Pyxis could be associated with the LMC \citep{2000PASP..112.1305P} or the ATLAS stream \citep{2014MNRAS.442L..85K}, but \citep{2017ApJ...840...30F} measured its proper motions with HST and concluded that it's likely associated with an unknown dwarf galaxy. While our hypothesis is that Cocytos is a disrupted globular cluster and not a dwarf galaxy, we have started a high-resolution spectroscopic follow-up with Magellan/MIKE for Cocytos that will measure its metallicity dispersion with higher accuracy and detailed chemical abundances for its members, providing more clues on its nature.

\section{Summary}\label{sec:summary}

The DESI Milky Way Survey offers an unprecedented opportunity to discover previously unidentified streams and dwarf galaxies based on radial velocities, as it provides the largest number of distant ($>$30~kpc) stars, surpassing previous spectroscopic surveys. To search for new streams, we applied spatial and kinematic masks, effectively masking two well-known streams: Sagittarius and Orphan. We then employed the OPTICS clustering algorithm to identify coherent groups of $>3$ giant stars, recovering 41 clusters with velocity dispersion $<$25~km/s, the majority of which corresponded to parts of previously known streams, globular clusters, and dwarf galaxies. However, a particularly intriguing structure emerged from this analysis, which lies near the Cocytos stream. Cocytos, which has not been kinematically confirmed thus far. We have kinematically selected more members using the DESI year 3 data, selected more members in Gaia and followed them up with Magellan/MagE.
We summarize our findings as follows:
\begin{enumerate}
\item Our blind search in Gaia uncovers a structure of \CocNGaiaCandRnd{} comoving stars at a distance of $\approx$ \CocOrbitDistRnd ~kpc.

 \item DESI spectroscopy and follow-up observations with Magellan/MagE reveal \CocNUniqueRnd{} confident members of the stellar stream that spans $\approx$~ 80$^\circ$.

\item This kinematically discovered stream spatially overlaps with the Cocytos stream discovered in photometry by \cite{2009ApJ...693.1118G} and is consistent with a very recent discovery by \cite{2025arXiv250322979T}.

\item With Gaia-selected members, we estimate a luminosity of M$_{\rm V} = \CocMvFmt$. With DESI and Magellan/MagE spectroscopic members, we measured a median [Fe/H]$=\CocFeHRnd$.

\item Using DESI and MagE spectroscopy, we estimate a velocity dispersion of $\CocSigmaVFmt$ km s$^{-1}$ and a metallicity dispersion of $\approx$\CocFeHDispRnd~dex.

\item We compared the orbital properties of the stream to other nearby structures, including Gaia--Enceladus, the Virgo Overdensities, and the 300-S stream. Cocytos has a radial, eccentric orbit, with radial actions and total orbital energies similar to 300-S. Based on this orbit and its [Fe/H], we conclude that Cocytos is likely a GC accreted with the infall of GSE, similar to the interpretation of the 300-S stream. This scenario could, in part, explain its relatively high-velocity dispersion. We also further confirm the association with the globular cluster Pyxis, which has the same [Fe/H], $J_R$, $L_z$, and total orbital energy as Cocytos. We are not sensitive to most of the structure's main-sequence turnoff and main-sequence population, preventing us from more detailed modeling of its age and star formation history.

\end{enumerate}

Detailed chemical abundance measurements and dynamical modeling will be needed to confirm our hypotheses --- that Cocytos had a globular cluster progenitor and that it was formed in GSE before its accretion into the Milky Way.

\begin{acknowledgments}
We thank Charlie Conroy, Nelson Caldwell, Ana Bonaca, Catherine Zucker, Suchetha Cooray, Carles Garcia Palau and Jesse Han for valuable discussions and comments that partially motivated and improved this work. We thank the staff at Las Campanas Observatory --- including Yuri Beletsky, Carla Fuentes, Jorge Araya, Hugo Rivera, Alberto Past\'en, Mauricio Mart\'inez, Roger Leiton, Mat\'ias D\'iaz, Carlos Contreras, and Gabriel Prieto --- for their invaluable assistance. Some computations in this paper were run on the FASRC Cannon cluster supported by the FAS Division of Science Research Computing Group at Harvard University. C. A. was funded through the Stanford Science Fellowship. 
T.S.L. acknowledges financial support from Natural Sciences and Engineering Research Council of Canada (NSERC) through grant RGPIN-2022-04794. S.K. acknowledges support from the Science \& Technology Facilities Council (STFC) grant ST\/Y001001\/1. 
This work received support from the U.S. Department of Energy (DOE) under contract number DE-AC02-76SF00515 to SLAC National Accelerator Laboratory, and under Contract No. DE–AC02–05CH11231, and by the National Energy Research Scientific Computing Center, a DOE Office of Science User Facility under the same contract. Additional support for DESI was provided by the U.S. National Science Foundation (NSF), Division of Astronomical Sciences under Contract No. AST-0950945 to the NSF’s National Optical-Infrared Astronomy Research Laboratory; the Science and Technology Facilities Council of the United Kingdom; the Gordon and Betty Moore Foundation; the Heising-Simons Foundation; the French Alternative Energies and Atomic Energy Commission (CEA); the National Council of Humanities, Science and Technology of Mexico (CONAHCYT); the Ministry of Science, Innovation and Universities of Spain (MICIU\/AEI\/10.13039\/501100011033), and by the DESI Member Institutions: \url{https://www.desi.lbl.gov/collaborating-institutions}. Any opinions, findings, and conclusions or recommendations expressed in this material are those of the author(s) and do not necessarily reflect the views of the U. S. National Science Foundation, the U. S. Department of Energy, or any of the listed funding agencies. The authors are honored to be permitted to conduct scientific research on Iolkam Du’ag (Kitt Peak), a mountain with particular significance to the Tohono O’odham Nation. This work has made use of data from the European Space Agency (ESA) mission Gaia (\url{https://www.cosmos.esa.int/gaia}), processed by the Gaia Data Processing and Analysis Consortium (DPAC, \url{https://www.cosmos.esa.int/web/gaia/dpac/consortium}). Funding for the DPAC has been provided by national institutions, in particular, the institutions participating in the Gaia Multilateral Agreement. This research has made extensive use of NASA's Astrophysics Data System Bibliographic Services. For the purpose of open access, the author has applied a Creative Commons Attribution (CC BY) license to any Author Accepted Manuscript version arising from this submission.
\end{acknowledgments}

\software{\texttt{numpy} \citep{Harris2020}, 
\texttt{scipy} \citep{Virtanen2020}, 
\texttt{matplotlib} \citep{Hunter2007}, 
\texttt{astropy} \citep{astropy2013,astropy2018,astropy2022},
\texttt{gala} \citep{gala,adrian_price_whelan_2020_4159870},
\texttt{MINESweeper} \citep{Cargile2020}
}

\facilities{Magellan:Baade (MagE), Gaia, Sloan, PS1, 2MASS, WISE}

\section{Appendix}

\begin{figure*}
    \centering
    \includegraphics[width= \textwidth]{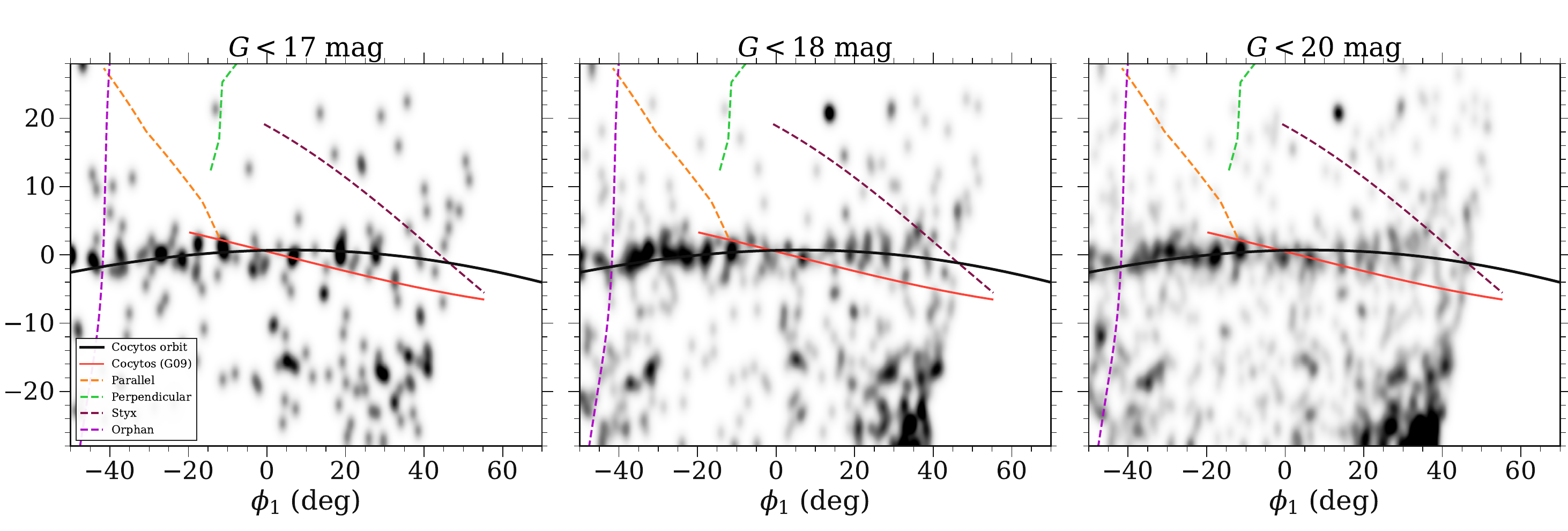}
    \caption{Selection of Cocytos as a function of Gaia depth at $G =17, 18, 20$ mag. In each panel, we applied our proper motion and isochrone selection. All panels show the density of stars in Gaia smoothed with a bandwidth of 1$^\circ$. Nearby streams are also shown with lines, and the stream's orbit, which nicely traces the overdensity, is shown as a solid black line. The stream track by \cite{2009ApJ...693.1118G} overlaps but does not quite match our computed orbit. We see significant contamination towards the Galactic plane, but overall, the stream is detectable with its width visually consistent with our estimate of $\approx$\CocWidthDegRnd$^\circ$ on the sky.}
    \label{fig:contamination}
\end{figure*}

Figure \ref{fig:contamination} shows the Gaia selection of Cocytos as a function of depth, showcasing an overlap with other nearby streams and structures in this region of the sky.

\bibliography{bibl}

\end{document}